\documentclass[11pt,a4paper]{article}

\usepackage[T1]{fontenc}
\usepackage[utf8]{inputenc}
\usepackage[a4paper,portrait,margin=1in]{geometry}
\usepackage{amsmath,amssymb,mathtools}
\usepackage{booktabs,array,tabularx,longtable,threeparttable,makecell,multirow}
\usepackage{graphicx}
\usepackage{float}
\usepackage{placeins}
\usepackage{caption}
\usepackage{subcaption}
\usepackage{xcolor}
\usepackage{hyperref}
\usepackage{xurl}
\usepackage{microtype}
\usepackage{enumitem}
\usepackage{listings}
\usepackage{fancyhdr}
\usepackage{authblk}
\usepackage{lastpage}
\usepackage{tikz}
\usetikzlibrary{arrows.meta,positioning,calc,fit,shapes.geometric}

\graphicspath{{figures/}}
\hypersetup{
  colorlinks=true,
  linkcolor=blue,
  citecolor=blue,
  urlcolor=blue,
  pdftitle={Ex Ante Estimation of Payable Relief and Compensation Timing for Supplier Selection},
  pdfauthor={Peplluis Esteva de la Rosa, Amogh Deshmukh}
}
\setlist{itemsep=0.25em,topsep=0.3em}
\newcommand{\EUR}{\ensuremath{\mathrm{EUR}}}
\newcommand{\PMR}{\mathrm{PMR}}
\newcommand{\CPM}{\mathrm{CPM}}
\newcommand{\CCA}{\mathrm{CCA}}
\newcommand{\CA}{\mathrm{CA}}
\newcommand{\ER}{\mathrm{ER}}

\newcommand{\ind}{\mathbf{1}}
\newcolumntype{Y}{>{\raggedright\arraybackslash}X}
\newcolumntype{L}[1]{>{\raggedright\arraybackslash}p{#1}}

\title{\textbf{Ex Ante Estimation of Payable Relief and Compensation Timing for Supplier Selection}}
\author[1]{Peplluis Esteva de la Rosa\thanks{Corresponding author: \href{mailto:joseplluis.delarosa@udg.edu}{joseplluis.delarosa@udg.edu}}}
\author[2]{Amogh Deshmukh}
\affil[1]{Universitat de Girona, Catalonia, EU}
\affil[2]{Woxsen University, Hyderabad, India; \href{mailto:amogh.deshmukh@woxsen.edu.in}{amogh.deshmukh@woxsen.edu.in}}
\date{September 7, 2026}

\begin{document}
\maketitle
\thispagestyle{plain}

\begin{abstract}
Supplier selection affects not only operating performance but also the payable network entered by a new obligation. This paper develops the Compensability Capacity Assessment (CCA), an ex ante buyer--supplier measure of expected gross payable relief and its likely timing. CPM provides the bounded structural kernel; concave CCA variants add bilateral invoice capacity. The measure is tested on 749,952 analytical invoices issued during 2012--2023, using frozen nine-month histories and future weekly, monthly, quarterly, semester, and annual windows. Cycle-restricted and path-enabled clearing are independent outcome-generating environments used to validate the measure, not technologies compared by this study.

Across seven fully observed quarters in 2022--2023, log-CCA has a median Spearman correlation of 0.638 with future integrated relief; persistent relations carry 93.5\% of relief, and the highest-scoring relation captures 92.5\% of buyer-specific best relief. Predictability remains positive from week to year, with quarterly recalibration providing the best operating balance between signal, coverage, and timeliness. A timing analysis shows that, across the two validation environments, 89--91\% of attributed relief occurs within seven days of invoice issue and 94--96\% within thirty days, on average about sixteen days before contractual maturity. Log-CCA correlates 0.564 with thirty-day relief and 0.566 with relief-days. Its highest quartile has a 96.5\% median probability of thirty-day compensation, compared with 57.6\% in the lowest quartile. Conditional waiting-time prediction is weaker, so CCA should rank timely compensation opportunity rather than forecast an exact payment date. The findings link supplier placement, financial circularity, and working-capital exposure while motivating deployment, causal testing, and quarterly drift monitoring.
\end{abstract}


\begin{quote}
\small
\textbf{Terminology and source note.} \emph{Payable relief}, \emph{compensation}, and \emph{debt relief} mean reduction of gross trade obligations through a governed settlement process, not legal forgiveness. \emph{Compensation timing} is the date on which an admissible operation extinguishes or transforms an attributed claim in the experimental ledger; for a generated settlement instruction it is not necessarily the date on which cash reaches the creditor. \emph{Financial circularity} refers to obligation-network structures that support such relief and is distinct from circular-economy material loops. The principal pseudonymized database is deposited as Esteva de la Rosa (2026) \cite{esteva2026data}. The analytical sequence contains 749,952 invoices issued during 2012--2023. Q4 2023 has observed follow-up only through 8 February 2024 and is marked as incomplete wherever it appears.
\end{quote}

\section{Introduction}

Supplier selection is a consequential production and sourcing decision because the chosen supplier affects cost, quality, delivery, continuity, flexibility, innovation, sustainability, and risk. Dickson's early study established that vendor selection is inherently multi-criteria rather than a price-only exercise \cite{dickson1966}. Weber, Current, and Benton later documented how formal criteria and quantitative methods became central to the field \cite{weber1991}. Contemporary frameworks extend the decision from qualification to order allocation under different purchasing strategies \cite{saputro2022}.

The financial consequence of supplier selection is usually represented through purchase price, payment terms, credit risk, or access to supply chain finance. That treatment remains largely dyadic. A buyer assesses the financial attributes of a contract or counterparty but rarely asks how a contemplated payable changes the wider directed network of obligations. Yet two commercially acceptable suppliers can occupy very different network positions. One may connect the buyer to reciprocal claims; another may provide access to a maturity-compatible downstream chain; a third may be structurally attractive but lack the transaction capacity required to activate that opportunity at meaningful scale.

This paper studies that overlooked decision dimension. It develops the Compensability Capacity Assessment, or CCA, as an ex ante measure of the payable relief associated with entering the invoice network through a specific supplier relation. The measure is supplementary rather than substitutive. It applies after conventional requirements concerning price, quality, delivery, sustainability, compliance, and resilience have been satisfied. Its purpose is to distinguish qualified alternatives by the amount of gross obligation that their placement may enable the system to remove before residual cash settlement or external funding is required.

The central construct is expected payable-mass reduction. The formulation separates the probability that a contemplated invoice participates in a maturity-admissible compensation event from the relief conditional on that event. The Circularity Propensity Measure, or CPM, is a bounded transformation of the expected-relief multiple. It preserves structural ordering while avoiding an unbounded scale. CCA then combines the structural signal with balanced bilateral invoice capacity. Because real invoice volumes are highly concentrated, the preferred operational form uses a concave logarithmic capacity adjustment.

Amount alone is not the entire production-economic outcome. A euro of relief generated near invoice issue can reduce gross settlement exposure for longer than the same euro generated shortly before maturity. Earlier compensation may also release attention, credit limits, and payment-processing capacity sooner, although those downstream effects depend on legal discharge and actual payment finality. The paper therefore extends the CCA evaluation from whether relief occurs and how much occurs to when the compensation or claim-transformation event occurs. It reports issue-to-event delay, days before contractual maturity, relief generated within seven, thirty, sixty, and ninety days, and a relief-days measure that weights attributable relief by the time remaining to maturity.

Cycle-restricted netting and path-enabled compensation are used only as independent outcome-generating environments. They are not the substantive focus of the paper and are not ranked as competing technologies here. Their structural differences make validation more demanding: a useful ex ante measure should remain related to relief across reciprocal-cycle and local-path opportunities. The policies are executed on separate residual ledgers. Their edge-attributed outcomes are added only to create an integrated opportunity label, not a jointly executable settlement total.

The empirical study uses the pseudonymized invoice archive deposited in Mendeley Data \cite{esteva2026data}. The analytical sequence comprises 749,952 invoices issued during 2012--2023 and is processed at source-record level with issue dates, due dates, stable identifiers, and residual balances. Scores are frozen from a trailing nine-month history before each target period. Future outcomes are then observed over weekly, monthly, quarterly, semester, and annual windows. The main repeated sequence covers Q1--Q4 2022 and Q1--Q4 2023, with the final quarter identified as right-censored because its follow-up ends on 8 February 2024.

The quarter-ahead results are positive but not invariant. Across the seven fully observed quarters, log-CCA Spearman correlation with integrated effective relief ranges from 0.592 to 0.663 and has a median of 0.638. Persistent relations represent only part of the future relational perimeter but carry a median 93.5\% of future relief. Within buyer choice sets, selecting the highest frozen log-CCA relation captures a median 92.5\% of buyer-specific best relief and materially outperforms the mean candidate. Q4 2022 is not treated as a singular success; it is interpreted only as part of an unusually reciprocal and cycle-saturated second half of 2022.

The timing results add a second dimension of predictive validity. Across the complete quarters, the path-enabled environment generates a median 89.4\% of attributable relief within seven days and 94.0\% within thirty days; the corresponding cycle-restricted shares are 90.8\% and 95.9\%. The PMR-weighted mean operation occurs approximately four days after invoice issue and fifteen to sixteen days before contractual maturity. Log-CCA has a median Spearman correlation of 0.564 with integrated thirty-day relief and 0.566 with integrated relief-days. Relations in its highest quartile have a 96.5\% median incidence of positive thirty-day compensation, compared with 57.6\% in the lowest quartile. By contrast, association with the exact waiting time among relations that eventually receive relief is modest. CCA is therefore better interpreted as ranking the likelihood and economic amount of timely compensation than as predicting an exact payment date.

The horizon comparison reveals a smoothing--coverage trade-off. Sixteen pre-specified weekly samples provide a high-frequency stress test; every calendar month in 2022--2023 supplies a monitoring panel; quarters provide the main operational evaluation; semesters represent portfolio planning; and annual windows represent strategic exposure. Median rank correlation rises from 0.526 weekly to 0.615 monthly, 0.638 quarterly, 0.640 semiannually, and 0.681 for the single complete annual origin, while relation persistence declines. Longer horizons reveal more eventual compensability but cover a smaller share of the future relational perimeter. This pattern supports quarterly recalibration, monthly drift monitoring, and weekly exception diagnostics rather than one timeless score.

We ask six linked questions. First, do CPM and the CCA variants calculated before a target period remain positively associated with future effective relief? Second, does the association repeat across every quarter of 2022 and 2023 rather than depend on one favorable window? Third, which capacity transformation best balances rank validity, magnitude alignment, and buyer-specific supplier selection? Fourth, do higher CCA values identify relief that is not only larger but also more likely to occur early relative to issue and maturity? Fifth, how do predictability, relation persistence, relief coverage, and buyer-level value change from week to month, quarter, semester, and year? Sixth, how should the measure be calibrated and governed when network structure, participation, and compensation timing drift?

This work makes eight contributions. First, it defines compensability as a production-economic consequence of supplier placement. Second, it distinguishes the monetary expected-relief estimand, CPM as its structural representation, and CCA as its commercially activated operational form. Third, it validates the measure repeatedly across eight quarter origins and multiple temporal horizons. Fourth, it tests six formula variants and identifies log-CCA as the strongest operational compromise rather than claiming universal dominance over CPM. Fifth, it evaluates the actual buyer-conditioned selection problem, not only a global correlation. Sixth, it adds a timing layer through short-horizon relief, maturity lead, and relief-days. Seventh, it uses two structurally distinct compensation environments as construct-coverage tests rather than as the research object being compared. Eighth, it translates the evidence into an implementation architecture and a broader research agenda spanning procurement, treasury, circular supply chains, public payment networks, temporal graph science, and socially governed digital infrastructure.

The remainder of the paper develops the state of the art and research gap, derives the measure and its timing extensions, reports the quarterly and multi-horizon evidence, and discusses scientific, managerial, circularity, and societal applications.

\section{State of the art and theoretical positioning}

\subsection{The evolution of supplier-selection research}

Supplier-selection research began by identifying the attributes used in vendor decisions. Dickson's survey showed that quality, delivery, performance history, warranties, capacity, price, technical capability, and financial position all influenced selection \cite{dickson1966}. The importance of that contribution lies not in the permanence of any one ranking but in the recognition that supplier choice is a structured multi-attribute decision.

Weber, Current, and Benton reviewed the next generation of work and found that cost, quality, and delivery remained dominant while mathematical programming and weighting methods became more common \cite{weber1991}. Their synthesis positioned supplier selection within operations research rather than treating it as an informal purchasing judgment. De Boer, Labro, and Morlacchi subsequently divided the process into problem definition, criteria formulation, qualification, and final choice \cite{deboer2001}. That staged view is important for us here because CCA is not intended to qualify an unacceptable supplier. It enters after minimum operational and strategic requirements have been met.

Ghodsypour and O'Brien integrated the analytic hierarchy process with linear programming to connect qualitative supplier evaluation with order allocation \cite{ghodsypour1998}. The model illustrates how a new criterion can affect not only ranking but also the quantity assigned to each supplier. Aissaoui, Haouari, and Hassini showed that supplier choice and order-lot sizing are often interdependent rather than separable problems \cite{aissaoui2007}. That interdependence supports a transaction-specific measure because the expected financial consequence depends on the amount placed with the supplier.

Ho, Xu, and Dey documented the wide adoption of multi-criteria decision-making methods for supplier evaluation \cite{ho2010}. Their review also showed that models frequently combine subjective and objective criteria. Chai, Liu, and Ngai found continued expansion toward optimization, fuzzy methods, data envelopment analysis, and hybrid intelligent techniques \cite{chai2013}. The methodological diversity shows that supplier-selection research is receptive to additional decision variables when those variables are theoretically grounded and empirically validated.

Saputro, Figueira, and Almada-Lobo emphasized that the appropriate supplier-selection model depends on purchasing strategy and decision context \cite{saputro2022}. A network-liquidity criterion should therefore not receive a universal fixed weight. Its relevance increases when the buyer participates in a compensation platform, when short-term funding is constrained, when invoice maturities overlap sufficiently, and when candidate suppliers differ materially in network position.

Green and circular procurement widened the criteria set beyond immediate economic performance. Govindan and coauthors synthesized methods that incorporate environmental performance in supplier evaluation \cite{govindan2015}. Bai, Zhu, and Sarkis introduced explicit circularity considerations into supplier selection \cite{bai2024}. These developments establish an important precedent: supplier choice can internalize system-level consequences that are not visible in unit price. The present paper follows that logic but applies it to the financial structure of trade obligations.

Most supplier models remain dyadic even when they are multi-criteria. They evaluate the focal supplier's attributes, certifications, risks, or performance. The network effect created by the buyer--supplier relation is usually represented indirectly, if at all. CCA adds a relational and topological criterion: the supplier is evaluated partly as an entry point into a wider payable network.

\subsection{Supply-network structure and network-aware operations}

Supply chains are not simple linear chains. Borgatti and Li argued that social network analysis provides concepts and tools for studying the pattern of interorganizational relationships in supply contexts \cite{borgatti2009}. Their contribution is methodological because it shifts attention from isolated firms to positions, ties, paths, brokerage, and local structure.

Choi and Kim developed a network perspective on structural embeddedness and supplier management \cite{choi2008}. Their analysis implies that the value and risk of a supplier relationship depend partly on the surrounding network rather than only on bilateral attributes. Kim, Choi, Yan, and Dooley applied social network analysis to the structural investigation of supply networks \cite{kim2011}. Their empirical approach demonstrated that network metrics can reveal operational properties that firm-level descriptions miss.

Network centrality, however, is not equivalent to compensability. Degree counts ties, betweenness identifies brokerage, and PageRank-like measures reflect recursive prominence. None directly states whether a proposed payable can enter a maturity-compatible compensation event. A high-degree supplier can be connected to invoices that mature too early or too late. A less central supplier can provide a single high-capacity route that generates substantial relief. CCA is therefore policy-conditioned and time-conditioned rather than a generic centrality measure.

Ivanov and Dolgui used the idea of intertwined supply networks to explain viability under severe disruption \cite{ivanov2020}. Their perspective reinforces the need to evaluate network-level consequences of operational decisions. We address a different outcome, but we share the premise that local sourcing choices can alter system-wide performance.

\subsection{Working capital and supply chain finance}

Pfohl and Gomm defined supply chain finance as the inter-company optimization of financing and the integration of financial flows with material and information flows \cite{pfohl2009}. That definition moved working-capital management beyond the boundary of a single balance sheet. Wuttke, Blome, and Henke provided empirical evidence on the coordination of financial flows and the organizational conditions of financial supply chain management \cite{wuttke2013}.

Hofmann and Kotzab argued for a supply-chain-oriented approach to working-capital management rather than isolated optimization by individual firms \cite{hofmann2010}. Their reasoning is particularly relevant because reducing one firm's cash conversion cycle can transfer pressure to another firm. A compensation mechanism differs from a unilateral extension of payment terms because it can reduce gross obligations without necessarily imposing the full burden on a weaker counterparty.

Randall and Farris linked cash-to-cash variables to supply chain financing decisions \cite{randall2009}. Their work highlights the production-economic relevance of timing and working-capital exposure. CCA contributes an upstream input to that analysis by estimating how much payable stock may disappear before the residual cash-to-cash requirement is financed.

Gelsomino and coauthors reviewed supply chain finance and found that the field is organized around financial instruments, enabling technologies, and interorganizational arrangements \cite{gelsomino2016}. Zhou, Chen, and Lee later described a wider financial ecology involving multiple actors and institutional relationships \cite{zhou2022}. These perspectives support a network view, but the dominant managerial object remains the financing of obligations rather than their endogenous extinction.

Factoring, reverse factoring, dynamic discounting, inventory finance, and trade credit can change who provides liquidity, when cash is advanced, or how financing costs are distributed. Compensation addresses a logically prior question. Before financing the gross stock, how much of that stock can be removed by matching obligations under a valid settlement rule? The answer changes the residual funding base.

Patra, Wankhede, and Agrawal concluded that circular-economy practices within supply chain finance remain an emerging area \cite{patra2024}. Gon\c{c}alves, de Carvalho, and Fiorini found that the direct financial consequences of circular practices were insufficiently developed in the literature \cite{goncalves2022}. CCA responds with a narrow and measurable link between circular structure and working-capital exposure.

\subsection{Network clearing, netting, and obligation compression}

Eisenberg and Noe demonstrated that payments in an interconnected liability system must be solved jointly because each firm's ability to pay depends on receipts from others \cite{eisenberg2001}. The broader implication is that the value and feasibility of an obligation cannot always be assessed independently of the surrounding network.

Rogers and Veraart showed how defaults, recovery, and rescue policies interact with interbank network structure \cite{rogers2013}. Their setting differs from trade invoices, but the study reinforces the principle that outcomes depend on topology and policy. Duffie and Zhu examined how netting design affects counterparty exposure and showed that institutional architecture can change the benefits of clearing \cite{duffie2011}. This policy dependence is directly relevant to the present distinction between cycle-restricted and path-enabled environments.

Verhoeff analyzed efficient settlement of multiple debts and made the computational structure of multilateral settlement explicit \cite{verhoeff2004}. Gavril\u{a} and Popa developed a graph algorithm for clearing obligations among companies and applied it in the Romanian institutional setting \cite{gavrila2021}. Their work supplies an important empirical foundation for treating invoices as a directed weighted network.

The clearing literature establishes that topology matters, but it rarely converts that insight into an ex ante procurement variable. Settlement research generally starts with a given obligation network and asks how to clear it. Supplier selection occurs one step earlier. It asks which new relation should be created when the buyer still has a choice. CCA translates the expected consequence of that network placement into a supplier-comparison signal.

The distinction between reciprocal and progressive structure also requires care. A directed cycle supports mutual netting because obligations return to the starting party. A directed path can support compression when a governed mechanism substitutes a shorter obligation for multiple intermediate obligations. These mechanisms reward different local motifs. The measure therefore separates mechanism-specific probability and conditional relief before combining them into an expected-relief estimand.

\subsection{Circular economy, circular supply chains, and measurement}

Geissdoerfer and coauthors described the circular economy as a sustainability paradigm while emphasizing that circularity and sustainability are not identical \cite{geissdoerfer2017}. Kirchherr, Reike, and Hekkert documented substantial variation among definitions of the circular economy \cite{kirchherr2017}. Their analysis warns against using the term circularity without specifying what circulates, at what level, and toward what objective.

Korhonen, Honkasalo, and Sepp\"al\"a examined conceptual limitations of the circular economy and emphasized the importance of system boundaries \cite{korhonen2018}. We have taken that warning . Financial circularity is not presented as a substitute for material circularity, environmental performance, product-life extension, or regenerative design.

Farooque and coauthors defined circular supply chain management as the integration of circular thinking into the management of supply chains and surrounding ecosystems \cite{farooque2019}. Montag synthesized circular supply-chain definitions and identified a broad research agenda involving closing, slowing, narrowing, intensifying, and dematerializing resource loops \cite{montag2023}. These contributions operate primarily in material, product, and energy domains.

Measurement studies show that circularity is difficult to operationalize. Saidani and coauthors developed a taxonomy of circular-economy indicators and demonstrated the diversity of purposes, levels, and calculation methods \cite{saidani2019}. Kristensen and Mosgaard reviewed micro-level circularity indicators and found important differences in scope and data requirements \cite{kristensen2020}. Vegter, van Hillegersberg, and Olthaar concluded that performance-measurement systems for circular supply chains remained underdeveloped \cite{vegter2021}. De Koning, Kassahun, and Tekinerdogan found that supply-chain circularity benchmarking still lacks stable consensus \cite{dekoning2024}.

CCA uses the word circularity in a deliberately restricted financial sense. The relevant resource is liquidity tied to gross obligations. The relevant loop is an obligation structure that permits claims to be mutually reduced or compressed under maturity and governance constraints. A network can also be financially compensable without containing a closed cycle when a path-enabled policy is available. For that reason, we distinguish reciprocal circularity from progressive compensability rather than equating all financial relief with literal graph cycles.

\subsection{Predictive decision support, temporal transport, and model drift}

An ex ante score must be judged differently from a descriptive index. A descriptive index can summarize the current network. A decision-support measure must retain useful ordering when applied to future or otherwise unseen outcomes. Same-period correlation is therefore necessary but insufficient.

Gama and coauthors reviewed concept-drift adaptation and showed that predictive relationships can change as the data-generating process evolves \cite{gama2014}. Invoice networks are especially exposed to drift because participation, transaction concentration, maturity practices, validation status, and settlement policy can change over time. A fixed structural formula may remain useful even when the mapping from that formula to euros requires recalibration.

This paper separates five forms of evidence. Annual direct validity tests association inside each period. Buyer-blocked cross-validation tests transfer to buyers excluded from model fitting. Forward calibration uses only earlier periods to estimate coefficients. Lagged relation transfer tests a prior-year score against next-year relief on the same relation. Strict at-issue validation freezes the score before invoice entry. The separation prevents a strong same-year coefficient from being misrepresented as a fully prospective forecast.

\subsection{Strategic capability and competitive advantage}

Kraljic argued that purchasing should be managed strategically rather than as a clerical function \cite{kraljic1983}. Barney's resource-based view explains why a valuable, scarce, inimitable, and non-substitutable resource can support sustained advantage \cite{barney1991}. Dyer and Singh located additional sources of advantage in interorganizational routines, relation-specific assets, complementary resources, and governance \cite{dyer1998}.

Teece, Pisano, and Shuen emphasized the ability to integrate, build, and reconfigure capabilities as environments change \cite{teece1997}. That perspective is useful because the formula itself is transparent and can be copied. The defensible capability lies in data coverage, maturity accuracy, partner participation, calibration, settlement governance, and the organizational routine that connects procurement decisions to treasury outcomes.

Financial circularity can therefore contribute to competitive advantage in three qualified ways. First, it can reduce avoidable gross settlement and associated funding exposure. Second, it can improve the timing and resilience of liquidity use. Third, repeated score-guided decisions can create learning about which categories, suppliers, and network positions generate relief. These benefits remain conditional on transaction costs, counterparty consent, legal enforceability, and the absence of adverse effects on price, quality, or supplier resilience.

\subsection{Research gap and positioning}

The literature streams converge on a clear gap. Supplier-selection research provides sophisticated ex ante methods but rarely represents the network liquidity consequence of placing a payable. Supply chain finance coordinates working capital but focuses primarily on financing or reallocating gross obligations. Clearing research demonstrates that network structure changes relief but normally assumes the obligation network is already given. Circularity research develops system-level indicators but does not measure the extinction of trade obligations. Predictive decision support warns about temporal transport but has not been applied systematically to a compensability measure.

\begin{table}[H]
\centering
\caption{Positioning of the article against adjacent literature streams.}
\label{tab:positioning}
\small
\begin{tabularx}{\textwidth}{L{0.17\textwidth}YYY}
\toprule
Literature stream & Established contribution & Unresolved issue & Our contribution \\
\midrule
Supplier selection & Qualification, ranking, and allocation across operational, economic, and sustainability criteria & Payable-network placement is absent from the normal criteria set & Adds a buyer-specific ex ante estimate of expected payable relief \\
Supply-network analysis & Shows that relational position and topology affect operational outcomes & Generic centrality does not encode maturity or settlement policy & Defines policy-conditioned and maturity-conditioned compensability \\
Supply chain finance & Coordinates financial flows and working capital across firms & Gross obligations are often financed before endogenous relief is estimated & Places compensation before or alongside residual external funding \\
Network clearing & Demonstrates that liability structure and policy alter settlement outcomes & The network is usually treated as given rather than shaped by procurement & Converts clearing opportunity into a pre-transaction supplier signal \\
Circularity measurement & Provides system-level concepts and indicator taxonomies & Material circularity does not measure obligation extinction & Defines a narrow financial-circularity construct grounded in payable relief \\
Predictive analytics & Distinguishes in-sample fit, transport, and drift & Limited application to supplier-level network-finance measures & Introduces a multi-layer annual, lagged, and at-issue validation protocol \\
Strategic capability & Explains advantage from data, routines, and relational governance & No operational construct links supplier placement to liquidity capability & Identifies conditions under which compensability intelligence can be valuable \\
\bottomrule
\end{tabularx}
\end{table}

\section{Hypothesis development and strategic proposition}

\subsection{Prospective validity}

A supplier relation creates value through compensation only when topology, bottleneck amounts, maturity intervals, and residual availability support an admissible operation. A score calculated before the target period should therefore remain positively associated with relief generated by subsequently issued invoices.

\textbf{H1.} \emph{CPM and the concavely capacity-adjusted CCA variants frozen before a target period are positively associated with effective relief attributed to invoices issued during that period.}

\subsection{Repeated quarterly validity}

A single successful quarter may reflect seasonality or one network regime. Repeating the same pre-specified design across Q1--Q4 2022 and Q1--Q4 2023 tests whether the relationship survives changes in participation, invoice concentration, maturity distribution, and reciprocal structure.

\textbf{H2.} \emph{Log-CCA is positively associated with effective relief in every quarter of 2022 and 2023, subject to the observed follow-up available for the final quarter.}

\subsection{Capacity transformation}

Raw CCA multiplies the structural score by an unbounded capacity ratio. Heavy-tailed commercial volume can therefore let a few very large suppliers dominate. Concave transformations should retain the activation role of capacity while limiting this leverage.

\textbf{H3.} \emph{Across quarter-ahead tests, log-CCA provides a more stable combination of rank correlation and magnitude alignment than raw CCA.}

\subsection{Supplier-selection benefit}

The intended decision is conditional on a buyer's feasible alternatives, not merely a global ordering of all relations. A useful measure should select relations that capture a high share of buyer-specific best relief and outperform the buyer-specific mean candidate.

\textbf{H4.} \emph{Selecting the highest-log-CCA relation among a buyer's observed eligible alternatives increases future relief relative to the mean candidate and captures a substantial share of the buyer-specific maximum.}

\subsection{Timely compensation opportunity}

Relief has greater working-capital relevance when it occurs early enough to reduce the duration for which gross obligations remain outstanding. A structurally and commercially favorable relation should therefore be associated not only with eventual relief but also with more relief generated within short post-issue horizons and with more relief-days before contractual maturity. Exact waiting time conditional on an event may remain difficult because execution order and residual competition add idiosyncratic variation.

\textbf{H5.} \emph{Log-CCA is positively associated with relief generated within thirty days and with relief-days before maturity, while its association with the exact conditional waiting time is weaker.}

\subsection{Horizon dependence}

Short windows contain fewer events and are more sensitive to exact issue timing and residual competition. Longer windows can average this noise but allow more new relations to appear after the score is frozen. Predictive signal and relational coverage therefore move on different margins.

\textbf{RQ1.} \emph{How do rank validity, log-magnitude alignment, relation persistence, relief coverage, buyer-specific decision value, and compensation timing change across one-week, one-month, one-quarter, one-semester, and one-year horizons?}

\subsection{Calibration and drift}

Even when the structural score transports, its monetary and temporal mapping can change with the network regime. A production estimator should therefore preserve the transparent score while recalibrating the conversion layer on prior completed windows.

\textbf{H6.} \emph{Sequential models trained only on completed prior quarters retain positive out-of-period predictive value, but their performance varies enough to justify recurrent recalibration and drift monitoring.}

\subsection{Strategic proposition}

The empirical hypotheses concern prediction, not causal firm performance. The strategic implication is conditional.

\textbf{Strategic proposition.} \emph{CCA can support a competitive liquidity capability when a firm or platform combines proprietary invoice-network visibility, reliable maturity and residual data, governed settlement participation, and procurement--treasury routines that competitors cannot readily reproduce.}

\section{Measure development: ex ante compensability and decision value}

\subsection{Decision setting and unit of analysis}

Let $G_t=(V,E_t,\mathcal T_t)$ be the directed invoice network observable at decision time $t$. Nodes are firms. A directed relation $A\to B$ means that buyer or debtor $A$ owes supplier or creditor $B$. The set $\mathcal T_t$ contains invoice tranches with issue date $\alpha_j$, due date $\tau_j$, original amount $w_j$, and current residual $r_j(t)$.

The contemplated decision is a new payable from buyer $A$ to a qualified supplier $B$ with amount $a$ and forecast horizon $H$. The target is not generic supplier attractiveness. It is the expected reduction in gross payable mass generated by placing that obligation at that network entry point.

A set of supporting tranches $S$ is maturity-admissible when their live intervals share a common date:
\begin{equation}
\max_{j\in S}\alpha_j \leq \min_{j\in S}\tau_j.
\label{eq:maturity}
\end{equation}
For an operation executed on date $d$, each supporting tranche must also satisfy $\alpha_j\leq d\leq\tau_j$ and $r_j(d)>0$. This condition prevents annual aggregation from creating a path or cycle whose component invoices were never simultaneously available.

\subsection{Effective payable-mass reduction}

Let $W_0$ be the gross residual amount before a settlement policy is applied, $W_1$ the gross residual of original obligations after the policy, and $I$ the amount of valid settlement instructions generated by the policy. Effective payable-mass reduction is
\begin{equation}
\PMR = W_0-(W_1+I).
\label{eq:pmr}
\end{equation}
The instruction term is essential. A policy that replaces two obligations by a direct obligation does not eliminate the full original gross amount. It removes only the intermediate payable mass after the generated instruction is restored to the post-policy stock.

At the relation level, effective relief is attributed to the entering buyer--supplier relation according to the deterministic operation log. Path-enabled and cycle-restricted policies are executed independently. Their attributed outcomes are denoted $R^{B}_{AB}$ and $R^{N}_{AB}$. The integrated opportunity target is
\begin{equation}
R^{I}_{AB}=R^{B}_{AB}+R^{N}_{AB}.
\label{eq:integrated}
\end{equation}
Equation~\eqref{eq:integrated} is a validation target for opportunity coverage. It is not a claim that the two separate policy maxima can be realized simultaneously on one residual ledger.

\subsection{Compensation timing and relief-days}

Let $f$ denote an attributed source-fragment contribution with invoice issue date $\alpha_f$, contractual due date $\tau_f$, operation date $d_f$, and effective-relief contribution $q_f$. Its issue-to-event delay and maturity lead are
\begin{equation}
\ell_f=d_f-\alpha_f,
\qquad
g_f=\min\{\max(\tau_f-d_f,0),365\}.
\label{eq:timing_terms}
\end{equation}
The one-year cap on $g_f$ prevents a very small number of implausibly distant due dates from dominating the time-weighted outcome. Removing the cap changes aggregate 2022--2023 relief-days by only 0.37\% in the path environment and 0.20\% in the cycle environment.

For relation $A\to B$, environment $m$, and short horizon $h$, timely relief and relief-days are
\begin{align}
R_{AB}^{m,\le h}&=\sum_{f\in\mathcal F_{AB}^{m}}q_f\ind\{\ell_f\le h\},
\label{eq:timely_relief}\\
RD_{AB}^{m}&=\sum_{f\in\mathcal F_{AB}^{m}}q_f g_f.
\label{eq:relief_days}
\end{align}
The empirical analysis uses $h\in\{7,30,60,90\}$ days. Equation~\eqref{eq:timely_relief} asks whether the score identifies relief that is realized soon after invoice issue. Equation~\eqref{eq:relief_days} rewards both amount and the period for which that amount is removed before contractual maturity.

These outcomes describe the timing of an algorithmically admissible compensation or claim-transformation event. They are not automatically equivalent to cash receipt. An accepted cycle reduction can extinguish the relevant claims at the event date under the governing legal arrangement. A non-bilateral path operation may instead create a settlement instruction whose payment and final discharge occur later. The analysis therefore supports claims about timely compensability and reduced gross-exposure duration, not an exact creditor cash-arrival date.

\subsection{Expected relief and mechanism decomposition}

Let $\mathcal M$ be the set of admissible compensation mechanisms. For mechanism $m$, let $P_{AB}^{m}(a,H)$ be the probability that the contemplated payable participates in a valid operation and let $\lambda_{AB}^{m}(a,H)$ be the conditional relief multiple. Expected relief is
\begin{equation}
\ER_{AB}(a,H)=a\sum_{m\in\mathcal M}P_{AB}^{m}(a,H)\lambda_{AB}^{m}(a,H).
\label{eq:expected_relief}
\end{equation}
The decomposition separates frequency from severity. A relation can have a low probability of activation but a high conditional multiple, or a high probability with modest conditional relief. The distinction matters for monitoring and governance because the two cases can have the same expected value but different operational risk.

\subsection{Cycle-restricted and path-enabled validation environments}

For a directed cycle $c$ with length $|c|$ and residual amounts $r_e$, the executable bottleneck is
\begin{equation}
\delta(c)=\min_{e\in c}r_e.
\end{equation}
A cycle step reduces every supporting obligation by $\delta(c)$ and produces gross relief
\begin{equation}
R^{N}(c)=|c|\delta(c).
\label{eq:cycle_relief}
\end{equation}
The cycle-restricted outcome environment enumerates exact elementary cycles through a common length cap of eight, matching the companion rolling-state study. Shorter caps are retained as robustness checks. The path-enabled environment uses local two-edge paths. These are outcome generators for validation of CCA, not technologies ranked by this work.

For a local path $A\to B\to C$, the matched amount is
\begin{equation}
s=\min\{r_{AB},r_{BC}\}.
\end{equation}
Both source obligations are reduced by $s$. If $A\neq C$, a direct instruction $A\to C$ of amount $s$ remains payable and effective relief equals $s$. If $A=C$, the operation is reciprocal cancellation and effective relief equals $2s$:
\begin{equation}
R^{B}(A,B,C;s)=\left(1+\ind\{A=C\}\right)s.
\label{eq:path_relief}
\end{equation}
The path policy rewards accessible two-edge structure and matched bottleneck amount. The cycle policy rewards reciprocal closure. Their different structural demands provide complementary tests of the measure.

\subsection{CPM as the bounded structural representation}

Define the compensability multiple
\begin{equation}
K_{AB}(a,H)=\frac{\ER_{AB}(a,H)}{a}.
\end{equation}
The Circularity Propensity Measure is
\begin{equation}
\CPM_{AB}(a,H)=100\left[1-\exp\left(-K_{AB}(a,H)\right)\right].
\label{eq:cpm}
\end{equation}
For fixed $a$, CPM is strictly increasing and concave in expected relief. It therefore preserves the structural ranking while compressing extreme multiples into a 0--100 interval. The inverse transformation is
\begin{equation}
\ER_{AB}(a,H)=-a\ln\left(1-\frac{\CPM_{AB}(a,H)}{100}\right).
\label{eq:cpm_inverse}
\end{equation}
CPM is not an independent sentiment score. It is the bounded structural representation of the monetary expected-relief proxy.

\subsection{Capacity adjustment and CCA}

Let $V_B^{\mathrm{sales}}$ be supplier $B$'s creditor-side invoice volume and $V_B^{\mathrm{purchases}}$ its debtor-side volume over the relevant historical window. Balanced bilateral capacity is
\begin{equation}
\CA_B(a)=\frac{\sqrt{V_B^{\mathrm{sales}}V_B^{\mathrm{purchases}}}}{a}.
\label{eq:capacity}
\end{equation}
The geometric mean rewards two-sided activity and limits the influence of a firm that is large on only one side of trade.

A general capacity-adjusted score is
\begin{equation}
\CCA^{g}_{AB}(a,H)=\CPM_{AB}(a,H)g\!\left(\CA_B(a)\right).
\label{eq:cca_general}
\end{equation}
Raw CCA uses $g(z)=z$. The preferred operational specification uses
\begin{equation}
g(z)=\ln(1+z).
\label{eq:log_capacity}
\end{equation}
The concave transform preserves complementarity between structure and capacity while limiting domination by extreme volume. We refer to this form as log-CCA.

Expected relief, CPM, and CCA have different units and roles. Expected relief is monetary. CPM is a bounded structural index. CCA is a capacity-adjusted prioritization feature. A CCA value should not be converted into euros without a separately estimated calibration model for the relevant period and market.

\subsection{Production-economic interpretation}

The measure links supplier placement to the residual gross settlement requirement. If a share $q_{AB}$ of expected relief would otherwise require external funding, the annual financing rate is $r$, and the expected funding duration avoided is $d$ days, a transparent scenario is
\begin{equation}
C_{AB}^{\mathrm{scenario}}=q_{AB}\ER_{AB}r\frac{d}{365}.
\label{eq:funding_scenario}
\end{equation}
Equation~\eqref{eq:funding_scenario} is not an observed causal saving in this study. It is a sensitivity translation. Platform fees, discounts, legal costs, implementation costs, default exposure, and changes in payment timing must be deducted before a net-benefit claim is made.

The measure can enter a conventional supplier scorecard after qualification. Its weight should depend on category strategy, funding conditions, confidence in the data, and access to governed compensation. It should not override a material price, quality, continuity, sustainability, or compliance disadvantage.

\begin{figure}[H]
\centering
\resizebox{0.98\textwidth}{!}{%
\begin{tikzpicture}[
  node distance=8mm and 9mm,
  box/.style={draw,rounded corners,align=center,minimum height=11mm,text width=31mm},
  benefit/.style={draw,rounded corners,align=center,minimum height=10mm,text width=29mm},
  arrow/.style={-{Latex[length=2.4mm]},thick}
]
\node[box] (decision) {Qualified supplier relation\\amount $a$, horizon $H$};
\node[box,right=of decision] (network) {Prior information\\amounts and maturities};
\node[box,right=of network] (estimate) {Expected relief\\$a\sum_m P^m\lambda^m$};
\node[box,right=of estimate] (score) {CPM and calibrated\\capacity adjustment};
\node[benefit,right=of score,yshift=15mm] (rank) {Supplier ranking};
\node[benefit,right=of score] (fund) {Funding-exposure\\scenario};
\node[benefit,right=of score,yshift=-15mm] (learn) {Outcome feedback\\and recalibration};
\draw[arrow] (decision) -- (network);
\draw[arrow] (network) -- (estimate);
\draw[arrow] (estimate) -- (score);
\draw[arrow] (score) -- (rank);
\draw[arrow] (score) -- (fund);
\draw[arrow] (score) -- (learn);
\end{tikzpicture}%
}
\caption{CCA converts a contemplated supplier relation into an ex ante production-economic signal. Settlement schemes provide empirical probability and magnitude environments; they are not the focal contribution.}
\label{fig:architecture}
\end{figure}
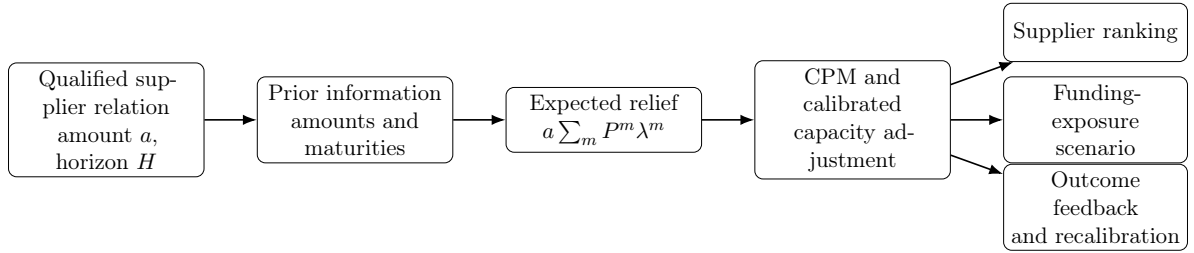

\subsection{Worked example}

Suppose buyer $A$ considers a new invoice of $a=\EUR\ 100{,}000$ to supplier $B$. The estimated cycle-event probability is 0.30 and the conditional cycle-relief multiple is 1.50. The estimated path-event probability is 0.60 and the conditional path-relief multiple is 0.75. Expected relief is
\[
\ER=100{,}000[(0.30)(1.50)+(0.60)(0.75)]=\EUR\ 90{,}000.
\]
The compensability multiple is 0.90 and CPM is $100(1-e^{-0.90})=59.34$. If the supplier's capacity ratio is 20, raw CCA is 1,186.8 and log-CCA is $59.34\ln(21)=180.7$. The example shows why the concave capacity form is preferable when invoice volumes are heavy-tailed. It also shows why the monetary estimate and operational score should remain distinct.

\section{Methodology}

\subsection{Research architecture}

The study uses a staged but unified temporal architecture. The score is calculated before a target issue window. The two compensation policies are then used independently to generate source-fragment-attributed relief outcomes on one continuous chronological state per policy. The main analysis asks whether the ex ante formula variants predict those outcomes. It does not use their aggregate policy difference as its dependent scientific claim.

The empirical layers are: (i) an annual bridge audit that verifies continuity and reconciles the implementation with the companion algorithm study; (ii) eight frozen quarter origins covering Q1--Q4 2022 and Q1--Q4 2023; (iii) formula-variant comparisons; (iv) buyer-specific supplier-selection tests; (v) multi-horizon comparisons from week to year; and (vi) sequential calibration using completed prior quarters only.

\subsection{Source archive and analytical population}

The principal pseudonymized research archive is deposited as \emph{Atomic Common-Day Invoice Clearing: Pseudonymized Invoice Records and Reproducibility Data, 2012--2023} \cite{esteva2026data}. The rolling execution stream contains 751,192 pseudonymized atomic records. Of these, 749,952 are analytical invoices issued from 2012 through 2023 and total \EUR\ 99.705 billion. A further 1,240 records issued from 1 January through 8 February 2024 support the final bridge but do not enter any 2012--2023 issue-cohort denominator. The accepted 2021 evidence is drawn from twelve monthly workbooks and connected to the harmonized adjacent years through documented crosswalks; unresolved identities remain distinct rather than being guessed.

Each atomic record contains a stable source identifier, debtor, creditor, issue date, due date, original amount, and residual amount. The aggregate graph is used for candidate discovery, but executable amount is always supplied by atomic records active on one common day. This distinction prevents annual or period aggregation from creating false overlap between invoices that never coexist.

\subsection{Common-day capacity and effective relief}

For edge $e=(u,v)$ on day $t$, let $c_e(t;s)$ be the sum of residual source amounts that have been issued and have not passed their due dates in state $s$. For a fixed path or circuit with edge set $F$, common-day capacity is
\begin{equation}
\delta_F(t;s)=\min_{e\in F}c_e(t;s),
\qquad
\delta_F^*(s)=\max_t\delta_F(t;s).
\label{eq:common_day_v4}
\end{equation}
The operation day is logged, and every consumed fragment must be active on that day. The effective outcome remains post-instruction payable-mass reduction from Eq.~\eqref{eq:pmr}. Non-bilateral path operations receive one unit of PMR per matched unit because a direct endpoint instruction remains; reciprocal paths receive two units; a length-$k$ cycle receives $k$ units.

\subsection{Rolling physical state and the two-month annual bridge}

The annual rule follows the supplied companion study \cite{companionbridge2026}. Each policy has one continuous physical state. For cohort $y$, records issued during $y$ enter chronologically. January and February invoices of $y+1$ are then introduced as the terminal bridge of cohort $y$. After that bridge, unresolved cohort-$y$ invoices are closed for cohort reporting, while only the residual balances of the new $y+1$ bridge invoices continue into the remainder of their own issue year.

For bridge invoice $i$ and policy $m$,
\begin{equation}
 x_i=c^{m,\mathrm{bridge}}_i+r^{m,\mathrm{carry}}_i,
\label{eq:bridge_identity_v4}
\end{equation}
where $x_i$ is original face value, $c^{m,\mathrm{bridge}}_i$ is physical consumption during the preceding cohort's bridge, and $r^{m,\mathrm{carry}}_i$ is the only amount available afterward. Consumed bridge amount is never restored. A global introduced-UID set and cumulative-consumption ledger reject duplicate introduction or consumption above face value.

Physical execution and issue-cohort attribution are distinct. Cycle PMR is attributed by the issue year of each consumed leg. Reciprocal-path PMR is also attributed by leg. The one unit of PMR from a non-bilateral path is split symmetrically across its two consumed source fragments. This preserves additivity across issue cohorts without changing physical state.

\subsection{Independent outcome environments}

The path-enabled and cycle-restricted policies see the same chronological arrivals but run on separate residual ledgers. The cycle environment enumerates all elementary circuits through length eight and executes feasible circuits under deterministic PMR-aligned ordering. The path environment applies deterministic local two-edge operations, with reciprocal candidates valued at $2q$ and non-bilateral candidates at $q$. Both use earliest-due, earliest-issue, and source-identifier ordering for fragment consumption.

The CCA analysis does not infer that one policy is preferable. Their role is analogous to two outcome definitions that expose different parts of the opportunity set. For relation $e$, the primary validation label is
\begin{equation}
R^{I}_e=R^{P}_e+R^{C}_e,
\end{equation}
where $R^{P}$ and $R^{C}$ are generated on separate ledgers. This integrated label measures coverage across validation environments and is never added to produce an executable portfolio total.

\subsection{Timing outcomes and shorter-term compensation}

The source-fragment contribution logs retain source issue date, source due date, operation date, relation, environment, and attributed PMR. For each target window the analysis calculates the share of PMR generated within 7, 30, 60, and 90 days of issue, the PMR-weighted issue-to-event delay, the PMR-weighted days remaining to maturity, and the relief-days measure in Eq.~\eqref{eq:relief_days}. Relation-level timely outcomes are merged with the score frozen at the origin.

Predictive validity is assessed in three complementary ways. First, the score is correlated with zero-inclusive PMR generated within thirty days. Second, it is correlated with relief-days, which combines relief amount and maturity lead. Third, among relations with positive relief, it is correlated with the negative PMR-weighted issue-to-event delay. The third test is deliberately conditional and answers the harder question of whether the score orders exact waiting time after an event is already known to occur.

Buyer-specific timing tests repeat the procurement decision for thirty-day relief and relief-days. Score-quartile analysis compares the probability of positive thirty-day compensation and the average event delay among positive relations. All timing summaries use the seven quarters with complete observed follow-up as the primary inference set; Q4 2023 is reported separately as right-censored.

\subsection{Frozen-origin score construction}

For each forecast origin $t_0$, the history network contains only invoices issued during the immediately preceding nine calendar months. All score variants are frozen before the first target invoice is observed. A relation must be present in the historical network to receive a focal relation score; such relations are called persistent. Supplier sales, purchases, degree, reciprocity, path reach, and bounded-cycle access are also computed from history only.

The fixed reference invoice is \EUR\ 50,000. Six variants are evaluated: CPM; raw CCA; log-CCA; square-root CCA; CCA with capacity capped at the historical 95th percentile; and leave-one-relation-out log-CCA, which subtracts the focal relation from supplier volume. The formulas and hyperparameters are not reselected after observing any target window.

\subsection{Forecast windows, follow-up, and outcome attribution}

For target issue interval $[s,e]$, a source-fragment contribution enters the outcome when its source invoice was issued inside $[s,e]$ and its execution date is no later than $e+2$ calendar months. Thus the target window is followed by a two-month outcome bridge. Contributions are read from the same rolling non-reuse stream. A source fragment used during this follow-up has already been physically consumed and cannot reappear in a later annual or quarterly state.

The quarter partitions are disjoint: Q1--Q4 2022 and Q1--Q4 2023. The monthly partitions are likewise disjoint within their panel. Week samples are alternative high-frequency diagnostics and are not summed with monthly or quarterly results. A contribution can legitimately be examined in more than one horizon experiment, just as one observation can enter several robustness specifications, but no totals are combined across overlapping horizon panels.

A window is complete only when its entire two-month bridge is observed. The archive supports complete bridges through Q3 2023. Q4 2023 ends on 31 December 2023 but is observed only through 8 February 2024, giving 39 follow-up days rather than the planned 60. It is reported in the quarter sequence with an explicit censoring marker and excluded from complete-only horizon summaries.

\subsection{Quarter, month, semester, year, and week designs}

Quarter origins occur on the final day preceding each quarter from 31 December 2021 through 30 September 2023. Month origins occur before every calendar month in 2022 and 2023. Semester origins precede January and July, and annual origins precede January.

Weekly analysis uses two dates per quarter chosen before outcome inspection: the first complete Monday--Sunday week and the Monday--Sunday week containing approximately the quarter midpoint. This produces sixteen samples. The purpose is to enrich time-scale comparison without treating every overlapping seven-day window as an independent observation or choosing unusually favorable weeks after seeing the outcomes.

\subsection{Evaluation metrics}

Spearman correlation evaluates relation ranking. Log-Pearson is the Pearson correlation between $\log(1+\mathrm{score})$ and $\log(1+R)$ and evaluates heavy-tail-adjusted magnitude alignment. Top-decile capture is the share of total relief concentrated in the highest-scoring ten percent of persistent relations.

The buyer-specific test retains buyers with at least two persistent candidate relations and at least one positive outcome. Each metric selects the highest-scoring relation. Best-relief capture is selected aggregate relief divided by the sum of buyer-specific maxima. Uplift compares selected relief with the sum of buyer-specific mean-candidate relief. Exact-best hit rate is the share of buyers for which an exact maximum is selected. Positive-selection rate records whether the selected relation receives positive relief.

For quarters, uncertainty is summarized with 500 buyer-cluster bootstrap resamples. The resampling unit is buyer, preserving relations within buyer. These intervals are dependence-aware stability diagnostics for the observed network, not IID population confidence intervals.

\subsection{Sequential calibration}

A two-part hurdle model is estimated sequentially. Logistic regression estimates positive relief; ridge regression estimates $\log(1+R)$ among positive observations. Each quarter is predicted using only completed earlier quarters. The full-control specification contains target amount, supplier sales and purchases, in-degree, out-degree, reciprocity, path reach, cycle count, and log-CCA. Standardization and regularization are fixed across origins.

This model is not a replacement for the transparent score. It tests whether a recalibrated monetary layer can improve magnitude prediction and buyer-specific selection while preserving the fixed underlying construct.

\subsection{Audit and reconciliation}

Independent replay checks source existence, edge consistency, operation-day activity, fragment sums, instruction mass, PMR identities, nonnegative residuals, and firm-level net-position conservation. A global audit spans all annual phases and bridge boundaries. The corrected cycle engine reproduces every annual value to less than 0.00005 percentage points. The independently coded path engine differs by at most 0.371 percentage points, concentrated in 2022, on average lower than 0.1, because local deterministic ordering and tie resolution are not fully unique across implementations. The CCA results use one internally consistent path stream, and this difference is disclosed rather than hidden.

The continuous execution introduces 751,192 UIDs exactly once. The cycle stream contains 67,425 operations and the path stream 157,529 operations. Both report zero duplicate introductions, zero overconsumed records, zero negative residuals, and zero temporal violations. Physical PMR equals source-fragment-attributed PMR exactly in both streams.

Table~\ref{tab:boundary_nonreuse} documents selected annual boundaries. For each January--February pool, original mass equals preceding-cohort consumption plus carried residual to the cent. These checks are part of the outcome-construction methodology rather than hypotheses about CCA.

\begin{table}[H]
\centering
\caption{Selected boundary non-reuse audit. Amounts are EUR billions.}
\label{tab:boundary_nonreuse}
\small
\begin{tabular}{rrrrrrr}
\toprule
Bridge into & Records & Original & Cycle consumed & Cycle carried & Path consumed & Path carried \\
\midrule
2021 & 22,779 & 2.132 & 0.854 & 1.278 & 1.202 & 0.930 \\
2022 & 12,958 & 6.061 & 3.029 & 3.032 & 4.493 & 1.568 \\
2023 & 14,265 & 5.011 & 1.503 & 3.509 & 2.479 & 2.532 \\
2024 & 1,240 & 0.630 & 0.323 & 0.308 & 0.343 & 0.287 \\
\bottomrule

\end{tabular}
\begin{minipage}{0.94\textwidth}\footnotesize
For each policy and row, original mass equals consumed mass plus carried residual. The final row uses the observed 1 January--8 February 2024 follow-up.
\end{minipage}
\end{table}

The annual stream is also reconciled with the independent companion implementation. Table~\ref{tab:annual_bridge} is retained solely as a methodological audit of the labels later used for CCA validation.

\begin{table}[H]
\centering
\caption{Rolling annual-state audit used to generate the predictive labels.}
\label{tab:annual_bridge}
\small
\resizebox{\textwidth}{!}{%
\begin{tabular}{rrrrrrr}
\toprule
Issue year & Mass (bn) & Follow-up days & Cycle PMR & Path PMR & Cycle post-year increment (m) & Path post-year increment (m) \\
\midrule
2012 & 0.069 & 59 & 1.341\% & 0.911\% & 0.0 & 0.0 \\
2013 & 0.119 & 59 & 0.038\% & 0.036\% & 0.0 & 0.0 \\
2014 & 0.197 & 59 & 0.226\% & 0.793\% & 0.0 & 0.0 \\
2015 & 0.538 & 60 & 0.254\% & 0.355\% & 0.2 & 0.2 \\
2016 & 0.679 & 59 & 1.202\% & 1.307\% & 0.0 & 0.0 \\
2017 & 0.257 & 59 & 3.813\% & 4.344\% & 0.0 & 0.3 \\
2018 & 0.460 & 59 & 3.453\% & 5.060\% & 0.7 & 0.7 \\
2019 & 4.424 & 60 & 14.140\% & 15.905\% & 134.0 & 101.0 \\
2020 & 14.020 & 59 & 42.790\% & 47.596\% & 158.5 & 148.6 \\
2021 & 15.153 & 59 & 36.730\% & 43.497\% & 77.2 & 72.5 \\
2022 & 41.576 & 59 & 52.795\% & 56.826\% & 434.5 & 443.2 \\
2023 & 22.212 & 39 & 40.705\% & 47.731\% & 144.7 & 148.1 \\
\bottomrule

\end{tabular}}
\begin{minipage}{0.94\textwidth}\footnotesize
Amounts in the final two columns are issue-cohort-attributed EUR millions generated during the prescribed post-year observation interval. The two policy columns document independent outcome streams; they are not interpreted as a CCA result or as a technology ranking.
\end{minipage}
\end{table}

\section{Results}

\subsection{Quarter-ahead association across Q1--Q4 2022 and 2023}

Table~\ref{tab:quarterly_direct} reports the principal frozen-origin result. Log-CCA is positively related to integrated effective relief in every quarter. Among the seven fully observed quarters, Spearman correlation ranges from 0.592 to 0.663 and has a median of 0.638; log-Pearson ranges from 0.596 to 0.656 with a median of 0.632. Q4 2023 yields 0.529 Spearman and 0.491 log-Pearson under the available follow-up through 8 February 2024 and is not included in the complete-only median.

\begin{table}[H]
\centering
\caption{Quarter-ahead validity of the principal formula variants.}
\label{tab:quarterly_direct}
\small
\resizebox{\textwidth}{!}{%
\begin{tabular}{lrrrrrr}
\toprule
Target quarter & CPM $\rho$ & Raw CCA $\rho$ & Log-CCA $\rho$ & Log-CCA log-$r$ & Positive rate & Top-decile capture \\
\midrule
Q1 2022 & 0.563 & 0.574 & 0.617 & 0.596 & 79.9\% & 78.4\% \\
Q2 2022 & 0.586 & 0.606 & 0.645 & 0.632 & 84.7\% & 64.5\% \\
Q3 2022 & 0.601 & 0.583 & 0.638 & 0.648 & 82.0\% & 84.4\% \\
Q4 2022 & 0.642 & 0.614 & 0.663 & 0.656 & 79.8\% & 82.6\% \\
Q1 2023 & 0.614 & 0.581 & 0.647 & 0.632 & 77.7\% & 63.2\% \\
Q2 2023 & 0.611 & 0.468 & 0.597 & 0.602 & 77.7\% & 53.5\% \\
Q3 2023 & 0.589 & 0.467 & 0.592 & 0.605 & 82.8\% & 60.0\% \\
Q4 2023$^{a}$ & 0.515 & 0.408 & 0.529 & 0.491 & 78.5\% & 58.1\% \\
\bottomrule

\end{tabular}}
\begin{minipage}{0.94\textwidth}\footnotesize
$\rho$ is Spearman correlation. Positive rate and top-decile capture refer to the integrated outcome on persistent relations. $^{a}$Q4 2023 is observed only through 8 February 2024.
\end{minipage}
\end{table}

\begin{figure}[H]
\centering
\includegraphics[width=0.90\textwidth]{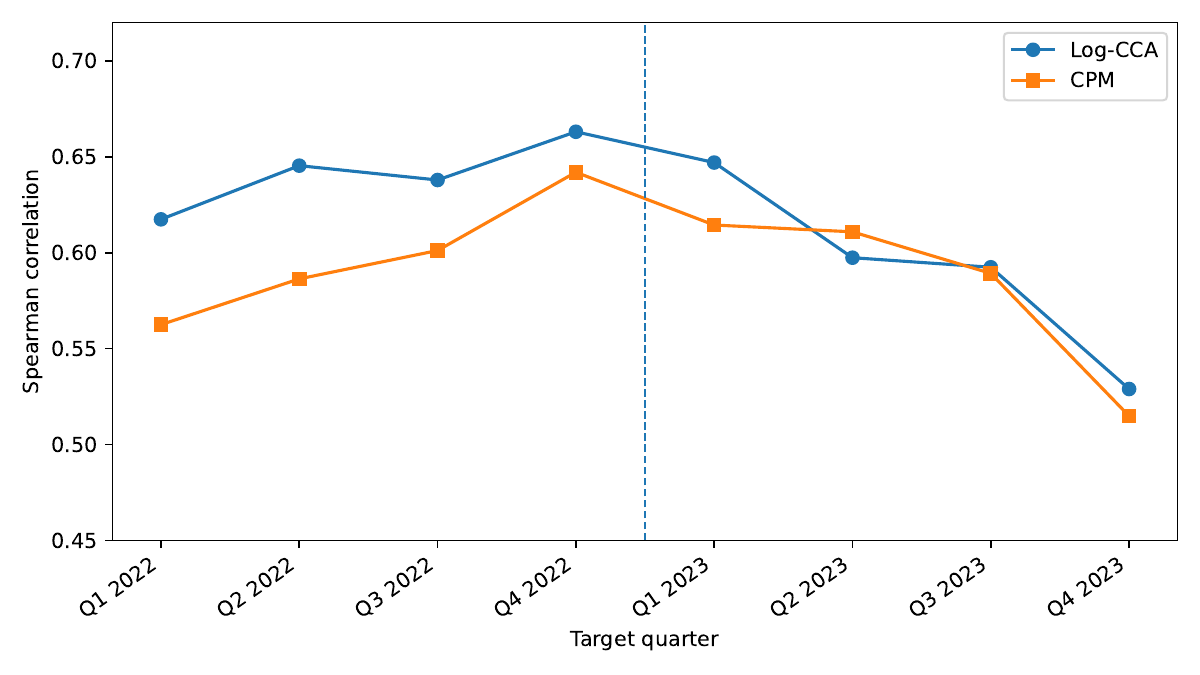}
\caption{Frozen log-CCA and future integrated relief by target quarter. Q4 2023 is right-censored because its follow-up is incomplete.}
\label{fig:quarterly_spearman}
\end{figure}

Buyer-cluster resampling confirms that the association is not carried by a single buyer. The 95\% bootstrap ranges are [0.578, 0.651], [0.607, 0.682], [0.605, 0.668], and [0.628, 0.699] for Q1--Q4 2022. They are [0.617, 0.679], [0.558, 0.637], and [0.556, 0.636] for the complete Q1--Q3 2023 windows. The partial Q4 2023 range is [0.456, 0.591]. All remain above zero.

These results support H1 and H2. They also change the interpretation of the earlier Q4 result. Q4 2022 is not unique because CCA works only there; it is one of eight positive quarter tests. Its relevance is structural: it belongs to the unusually reciprocal and cycle-saturated second half of 2022.

\subsection{Timely compensation and reduction of outstanding duration}

The amount results answer whether CCA identifies future relief. Table~\ref{tab:timing_descriptive} adds the temporal question: how soon is the attributed compensation or claim transformation generated? Across the seven fully observed quarters, both validation environments act quickly once an eligible source invoice enters the network. The PMR-weighted median event date is the issue day itself, reflecting the causal daily scheduler's ability to execute an already supported opportunity as soon as the new record becomes active. Mean delay remains approximately four days. The typical event occurs fifteen to sixteen days before contractual maturity.

\begin{table}[H]
\centering
\caption{Timing of attributed compensation events across fully observed quarters.}
\label{tab:timing_descriptive}
\small
\resizebox{\textwidth}{!}{%
\begin{tabular}{lrrrrrrr}
\toprule
Environment & Mean delay & Median delay & Mean days before due & Median before due & Within 7 days & Within 30 days & Within 60 days \\
\midrule
Path-enabled & 4.0 & 0 & 16.0 & 10 & 89.4\% & 94.0\% & 99.6\% \\
Cycle-restricted & 3.6 & 0 & 15.4 & 10 & 90.8\% & 95.9\% & 99.4\% \\
\bottomrule

\end{tabular}}
\begin{minipage}{0.95\textwidth}\footnotesize
Delay is measured from invoice issue to the operation that receives the source-fragment PMR attribution. Values are PMR-weighted medians across Q1 2022--Q3 2023. An event may extinguish a claim or create a settlement instruction; it is not necessarily a cash-receipt date.
\end{minipage}
\end{table}

The short-horizon shares are economically large. A median 89.4\% of path-attributed relief and 90.8\% of cycle-attributed relief is generated within seven days of invoice issue. The corresponding thirty-day shares are 94.0\% and 95.9\%, and more than 99\% is generated within sixty days. These percentages do not imply that all creditors receive cash within those periods. They show that the experimental ledger can identify and transform a large share of compensable gross obligation early in the invoice life cycle.

\begin{figure}[H]
\centering
\begin{subfigure}[t]{0.49\textwidth}
\centering
\includegraphics[width=\textwidth]{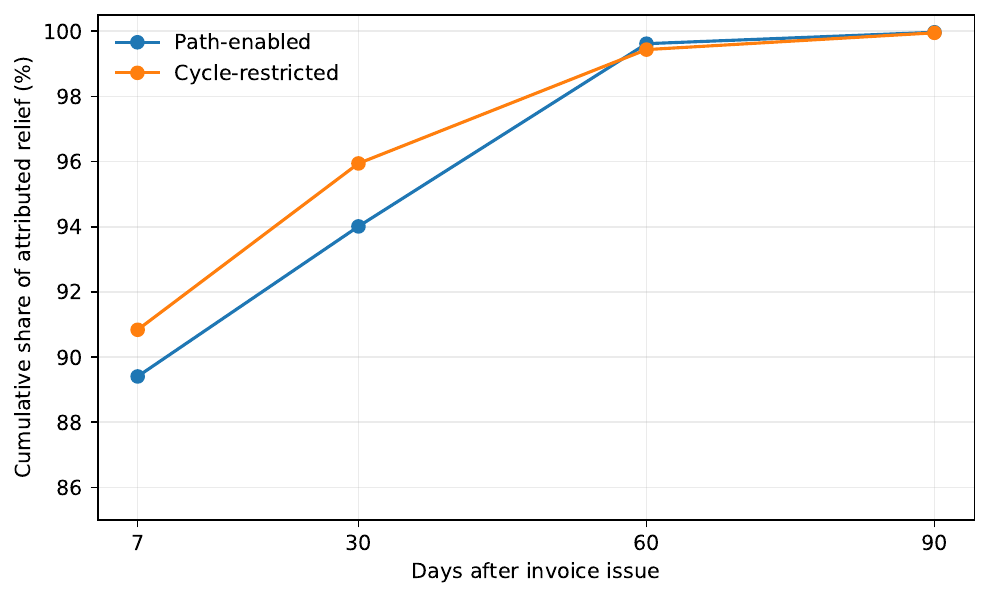}
\caption{Cumulative attributed relief by event delay.}
\end{subfigure}
\hfill
\begin{subfigure}[t]{0.49\textwidth}
\centering
\includegraphics[width=\textwidth]{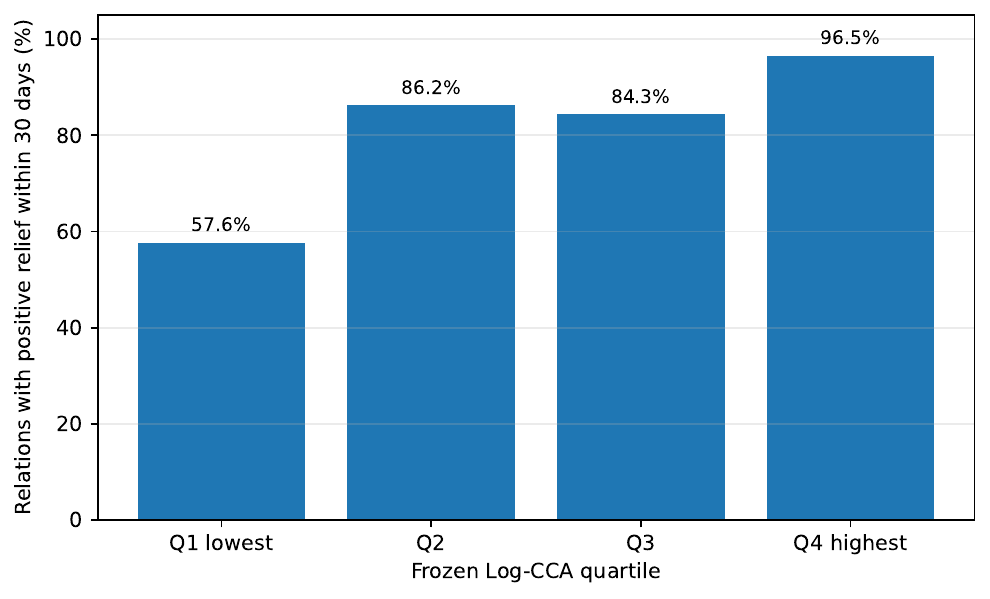}
\caption{Thirty-day incidence by score quartile.}
\end{subfigure}
\caption{Compensation timing and the frozen Log-CCA gradient. Complete quarters only.}
\label{fig:timing}
\end{figure}

The score predicts timely economic opportunity, although not every aspect of timing equally well. Table~\ref{tab:timing_predictive} reports the median quarter-level correlations for Log-CCA. In the integrated target, Log-CCA reaches 0.564 Spearman and 0.548 log-Pearson against relief generated within thirty days, and 0.566 and 0.538 against relief-days. Relations in the highest Log-CCA quartile have a median 96.5\% incidence of positive thirty-day relief, compared with 57.6\% in the lowest quartile. Their PMR-weighted event delay among positive relations is also lower at 3.3 days versus 8.8 days.

\begin{table}[H]
\centering
\caption{Predictive validity of Log-CCA for timely compensation outcomes.}
\label{tab:timing_predictive}
\small
\resizebox{\textwidth}{!}{%
\begin{tabular}{lrrrrr}
\toprule
Environment & \multicolumn{2}{c}{Relief within 30 days} & \multicolumn{2}{c}{Relief-days} & Shorter conditional term \\
\cmidrule(lr){2-3}\cmidrule(lr){4-5}
& Spearman & Log-Pearson & Spearman & Log-Pearson & Spearman \\
\midrule
Path-enabled & 0.560 & 0.540 & 0.565 & 0.531 & 0.241 \\
Cycle-restricted & 0.522 & 0.498 & 0.532 & 0.501 & 0.247 \\
Integrated opportunity & 0.564 & 0.548 & 0.566 & 0.538 & 0.209 \\
\bottomrule

\end{tabular}}
\begin{minipage}{0.95\textwidth}\footnotesize
Medians across Q1 2022--Q3 2023. ``Shorter conditional term'' correlates the score with the negative PMR-weighted delay only among relations with positive relief. Integrated values sum labels generated independently in the two validation environments and are not a joint settlement outcome.
\end{minipage}
\end{table}

The weaker conditional-term coefficient is an important boundary. Once relief is known to occur, Log-CCA has only a 0.209 median association with the exact relation-level delay in the integrated target. Residual competition, daily candidate order, and the arrival of supporting invoices determine the exact date. The measure should therefore be used to rank the likelihood and amount of timely compensation, not to promise a precise payment date.

The buyer-level timing experiment confirms decision relevance. Selecting the highest-Log-CCA relation captures a median 94.7\% of buyer-specific best relief generated within thirty days and produces a 307.6\% uplift over the mean candidate. For relief-days, the corresponding values are 91.2\% and 208.8\%.

\begin{table}[H]
\centering
\caption{Buyer-specific selection for timely compensation outcomes.}
\label{tab:timing_selection}
\small
\begin{tabular}{lrrrr}
\toprule
Outcome & Median buyers & Best captured & Uplift vs mean & Exact best \\
\midrule
Relief within 30 days & 183 & 94.7\% & 307.6\% & 58.6\% \\
Relief-days & 174 & 91.2\% & 208.8\% & 57.1\% \\
\bottomrule

\end{tabular}
\begin{minipage}{0.95\textwidth}\footnotesize
Medians across the seven fully observed quarters. Buyers require at least two persistent candidate relations and at least one positive outcome.
\end{minipage}
\end{table}

These results support H5. CCA's practical time contribution is not exact date forecasting. It is the identification of relations that are more likely to produce economically material compensation early enough to reduce the duration of outstanding gross obligations.

\subsection{Formula variants and the role of capacity}

Across complete quarters, median Spearman correlation is 0.638 for log-CCA, 0.610 for square-root CCA, 0.601 for CPM, 0.586 for capped CCA, 0.581 for raw CCA, and 0.385 for leave-one-relation-out CCA. Log-CCA also has the highest median log-Pearson correlation at 0.632. Raw CCA is not uniformly poor, but its unbounded capacity multiplier is less stable when supplier volumes are concentrated.

The result is more nuanced than saying that one formula dominates every objective. CPM captures a median 91.4\% of total integrated relief in its top decile, compared with 64.5\% for log-CCA, because CPM concentrates directly on structural opportunity. Log-CCA performs better as a balanced relation-level predictor and, as shown below, as a buyer-conditioned selection score. Capacity is therefore valuable when regularized and interpreted relative to the decision task.

\begin{figure}[H]
\centering
\includegraphics[width=0.88\textwidth]{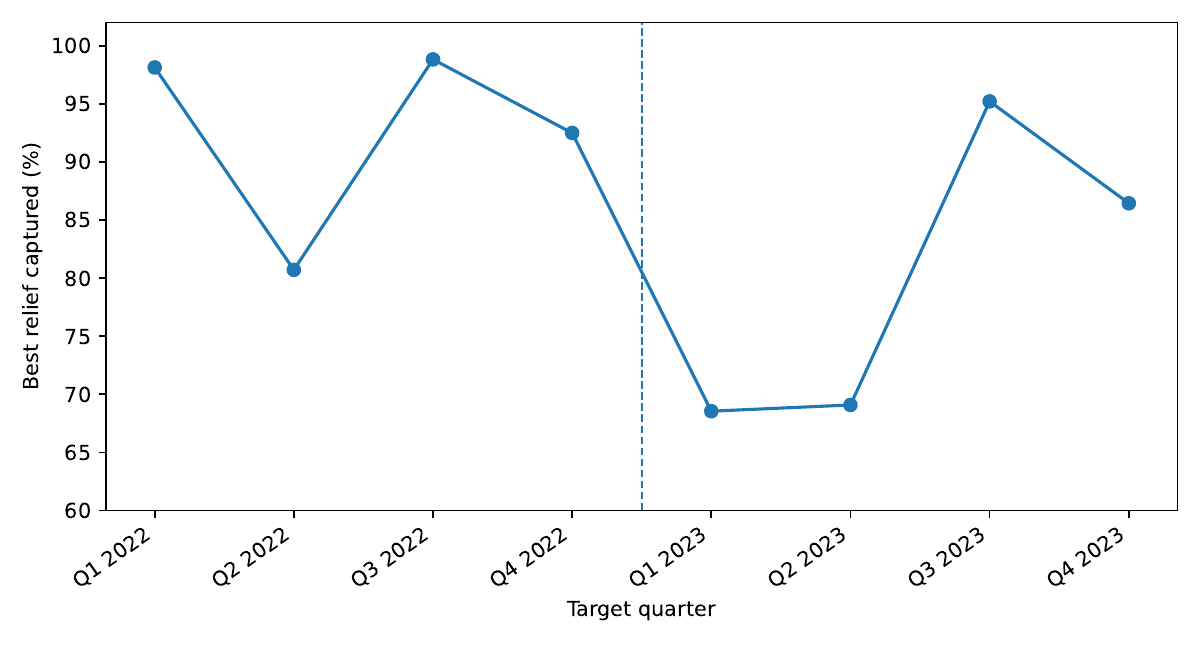}
\caption{Buyer-specific best-relief capture by frozen log-CCA across the eight quarter targets.}
\label{fig:selection_capture}
\end{figure}

The full quarter-by-quarter variant table appears in Appendix~\ref{app:variant_detail}. Its pattern supports H3: log-CCA is the most stable operational compromise, not a universal replacement for the structural CPM kernel.

\subsection{Independent path and cycle outcome checks}

Table~\ref{tab:quarterly_channels} reports log-CCA against each independently generated channel and the integrated target. All channel correlations are positive. Complete-quarter path correlations range from 0.511 to 0.563, while cycle correlations range from 0.472 to 0.578. The integrated target is consistently higher because it covers relations rewarded in either structural environment.

\begin{table}[H]
\centering
\caption{Log-CCA correlation across independent outcome-generating environments.}
\label{tab:quarterly_channels}
\small
\resizebox{\textwidth}{!}{%
\begin{tabular}{lrrrrrr}
\toprule
Quarter & Path $\rho$ & Path log-$r$ & Cycle $\rho$ & Cycle log-$r$ & Integrated $\rho$ & Integrated log-$r$ \\
\midrule
Q1 2022 & 0.535 & 0.433 & 0.514 & 0.506 & 0.617 & 0.596 \\
Q2 2022 & 0.563 & 0.445 & 0.534 & 0.540 & 0.645 & 0.632 \\
Q3 2022 & 0.533 & 0.449 & 0.561 & 0.586 & 0.638 & 0.648 \\
Q4 2022 & 0.557 & 0.475 & 0.578 & 0.579 & 0.663 & 0.656 \\
Q1 2023 & 0.542 & 0.438 & 0.568 & 0.569 & 0.647 & 0.632 \\
Q2 2023 & 0.511 & 0.425 & 0.509 & 0.511 & 0.597 & 0.602 \\
Q3 2023 & 0.541 & 0.442 & 0.472 & 0.458 & 0.592 & 0.605 \\
Q4 2023$^{a}$ & 0.542 & 0.433 & 0.410 & 0.399 & 0.529 & 0.491 \\
\bottomrule

\end{tabular}}
\begin{minipage}{0.94\textwidth}\footnotesize
The path and cycle policies are run independently. Integrated relief is a validation label and cannot be interpreted as the sum available from one joint execution. $^{a}$Incomplete observed follow-up.
\end{minipage}
\end{table}

This table is not an algorithm comparison. It shows that the same ex ante score remains informative under two different ways of realizing compensability. The cycle and path columns should therefore be read as complementary construct-validity checks.

\subsection{Persistent relations carry most future relief}

A frozen relation score exists only when the relation is already visible in the nine-month history. Table~\ref{tab:quarter_inventory} shows that persistence varies more than economic coverage. Across complete quarters, only 61.1--80.2\% of target relations are persistent, but they carry 73.8--95.9\% of integrated relief. The median relief share is 93.5\%.

\begin{table}[H]
\centering
\caption{Target-window scale and persistent-relation coverage.}
\label{tab:quarter_inventory}
\small
\resizebox{\textwidth}{!}{%
\begin{tabular}{lrrrrrrr}
\toprule
Quarter & Invoice rows & Target relations & Persistent & Rel. share & Mass share & Relief share & Follow-up days \\
\midrule
Q1 2022 & 20,010 & 2,485 & 1,829 & 73.6\% & 90.6\% & 95.1\% & 61 \\
Q2 2022 & 13,884 & 1,956 & 1,569 & 80.2\% & 84.5\% & 73.8\% & 61 \\
Q3 2022 & 23,149 & 3,135 & 1,917 & 61.1\% & 84.1\% & 89.7\% & 61 \\
Q4 2022 & 22,428 & 3,361 & 2,231 & 66.4\% & 94.6\% & 95.9\% & 59 \\
Q1 2023 & 22,549 & 3,055 & 2,097 & 68.6\% & 94.7\% & 95.5\% & 61 \\
Q2 2023 & 22,916 & 3,181 & 2,176 & 68.4\% & 92.1\% & 93.5\% & 61 \\
Q3 2023 & 22,303 & 3,219 & 2,049 & 63.7\% & 91.1\% & 92.8\% & 61 \\
Q4 2023$^{a}$ & 18,834 & 2,336 & 1,682 & 72.0\% & 86.5\% & 93.7\% & 39 \\
\bottomrule

\end{tabular}}
\begin{minipage}{0.94\textwidth}\footnotesize
Mass and relief shares are the proportions associated with relations visible at the forecast origin. $^{a}$Q4 2023 has partial observed follow-up.
\end{minipage}
\end{table}

This result explains why ex ante ranking can remain economically useful even when many future relations are new. Entry and exit occur predominantly around the relational perimeter, while a smaller persistent core carries most compensable value. Q2 2022 is the main exception: persistent relations account for 73.8\% of relief, still substantial but well below the other complete quarters.

\subsection{Buyer-specific supplier prioritization}

The buyer-level test is closer to the intended procurement use. Across complete quarters, choosing the highest-log-CCA relation captures a median 92.5\% of the sum of buyer-specific best outcomes and produces a median 348.4\% uplift over the sum of buyer-specific mean candidates. Exact-best hit rates are stable around 60--67\%, showing that very high euro capture does not require identifying the exact maximum for every buyer.

\begin{table}[H]
\centering
\caption{Buyer-specific selection using frozen log-CCA.}
\label{tab:quarter_selection}
\small
\begin{tabular}{lrrrrr}
\toprule
Quarter & Buyers & Best relief captured & Uplift vs mean & Exact best & Positive selected \\
\midrule
Q1 2022 & 160 & 98.1\% & 386.0\% & 61.9\% & 93.8\% \\
Q2 2022 & 146 & 80.7\% & 204.0\% & 63.0\% & 94.5\% \\
Q3 2022 & 179 & 98.8\% & 566.7\% & 60.9\% & 95.0\% \\
Q4 2022 & 200 & 92.5\% & 420.2\% & 64.5\% & 96.5\% \\
Q1 2023 & 177 & 68.5\% & 171.0\% & 66.7\% & 93.8\% \\
Q2 2023 & 191 & 69.1\% & 158.6\% & 60.2\% & 91.1\% \\
Q3 2023 & 186 & 95.2\% & 348.4\% & 62.4\% & 93.0\% \\
Q4 2023$^{a}$ & 153 & 86.4\% & 359.3\% & 62.7\% & 95.4\% \\
\bottomrule

\end{tabular}
\begin{minipage}{0.94\textwidth}\footnotesize
Buyers require at least two persistent relations and at least one positive candidate outcome. $^{a}$Incomplete observed follow-up.
\end{minipage}
\end{table}

The complete-quarter capture rate ranges from 68.5\% to 98.8\%. Lower capture in Q1 and Q2 2023 indicates that the global relationship alone does not guarantee equal decision quality in every buyer portfolio. Nevertheless, the selected relation remains markedly better than the mean candidate in every quarter. H4 is therefore supported with meaningful temporal heterogeneity.

\subsection{Horizon comparison from week to year}

Table~\ref{tab:horizon_summary} compares windows with complete observed follow-up. Median Spearman correlation rises from 0.526 at one week to 0.615 at one month, 0.638 at one quarter, 0.640 at one semester, and 0.681 at one year. This pattern is consistent with temporal aggregation smoothing sparse event timing and local residual-allocation noise.

\begin{table}[H]
\centering
\caption{Predictability and coverage across forecast horizons.}
\label{tab:horizon_summary}
\small
\resizebox{\textwidth}{!}{%
\begin{tabular}{lrrrrrrr}
\toprule
Horizon & Complete origins & Median $\rho$ & IQR $\rho$ & Median log-$r$ & Relation persistence & Relief coverage & Best capture \\
\midrule
One week & 16 & 0.526 & [0.489, 0.596] & 0.484 & 88.9\% & 96.6\% & 90.3\% \\
One month & 23 & 0.615 & [0.575, 0.628] & 0.577 & 81.9\% & 95.0\% & 96.7\% \\
One quarter & 7 & 0.638 & [0.607, 0.646] & 0.632 & 68.4\% & 93.5\% & 92.5\% \\
One semester & 3 & 0.640 & [0.633, 0.656] & 0.630 & 56.6\% & 85.1\% & 96.8\% \\
One year & 1 & 0.681 & [0.681, 0.681] & 0.654 & 50.3\% & 83.8\% & 97.9\% \\
\bottomrule

\end{tabular}}
\begin{minipage}{0.94\textwidth}\footnotesize
The annual row contains only the complete 2022 origin and should not be treated as a precise general estimate. All rows use log-CCA and integrated effective relief.
\end{minipage}
\end{table}

\begin{figure}[H]
\centering
\begin{subfigure}{0.49\textwidth}
\includegraphics[width=\textwidth]{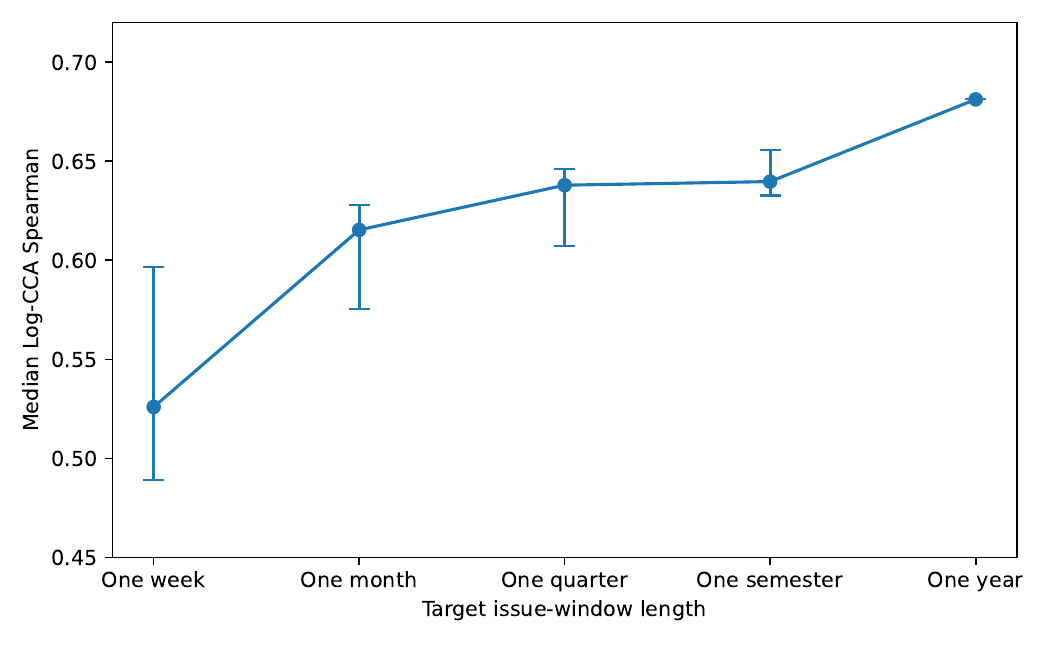}
\caption{Predictive signal.}
\end{subfigure}
\hfill
\begin{subfigure}{0.49\textwidth}
\includegraphics[width=\textwidth]{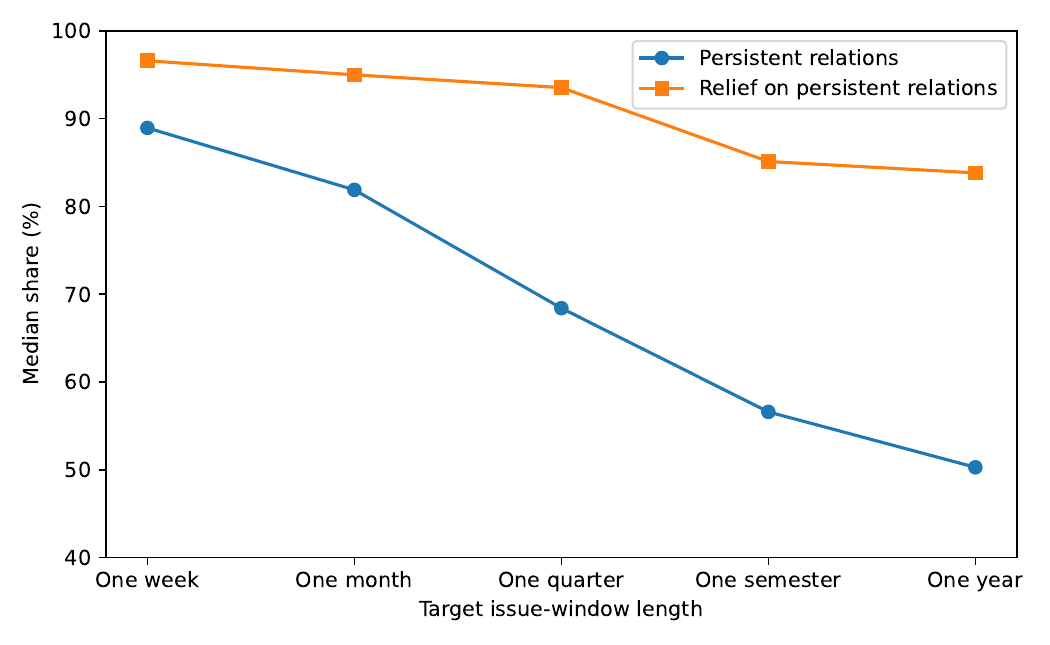}
\caption{Persistence and economic coverage.}
\end{subfigure}
\caption{Longer horizons smooth the outcome signal but reduce the proportion of future relations visible at the origin.}
\label{fig:horizon_tradeoff}
\end{figure}

The second margin is relation turnover. Median persistence falls from 88.9\% weekly to 81.9\% monthly, 68.4\% quarterly, 56.6\% semiannually, and 50.3\% annually. Relief coverage declines more slowly, from 96.6\% to 83.8\%. This supports the stable-core interpretation: many future relations are new, but the pre-existing core retains a disproportionate share of economically material relief.

The annual value must be interpreted cautiously because only the 2022 annual origin has complete observed follow-up. The 2023 annual value is a sensitivity with partial follow-up, not a second complete replication. RQ1 therefore supports a qualitative smoothing--coverage trade-off rather than a precise monotone law.

\subsection{Month-to-month evidence}

Monthly windows supply 23 complete origins and are useful for monitoring between quarterly recalibrations. Complete-month Spearman correlations range from 0.517 to 0.659, with a median of 0.615. The month sequence is positive throughout, but buyer-specific top-decile and selection outcomes are more volatile because monthly relief can be concentrated in a few relations.

\begin{figure}[H]
\centering
\includegraphics[width=0.92\textwidth]{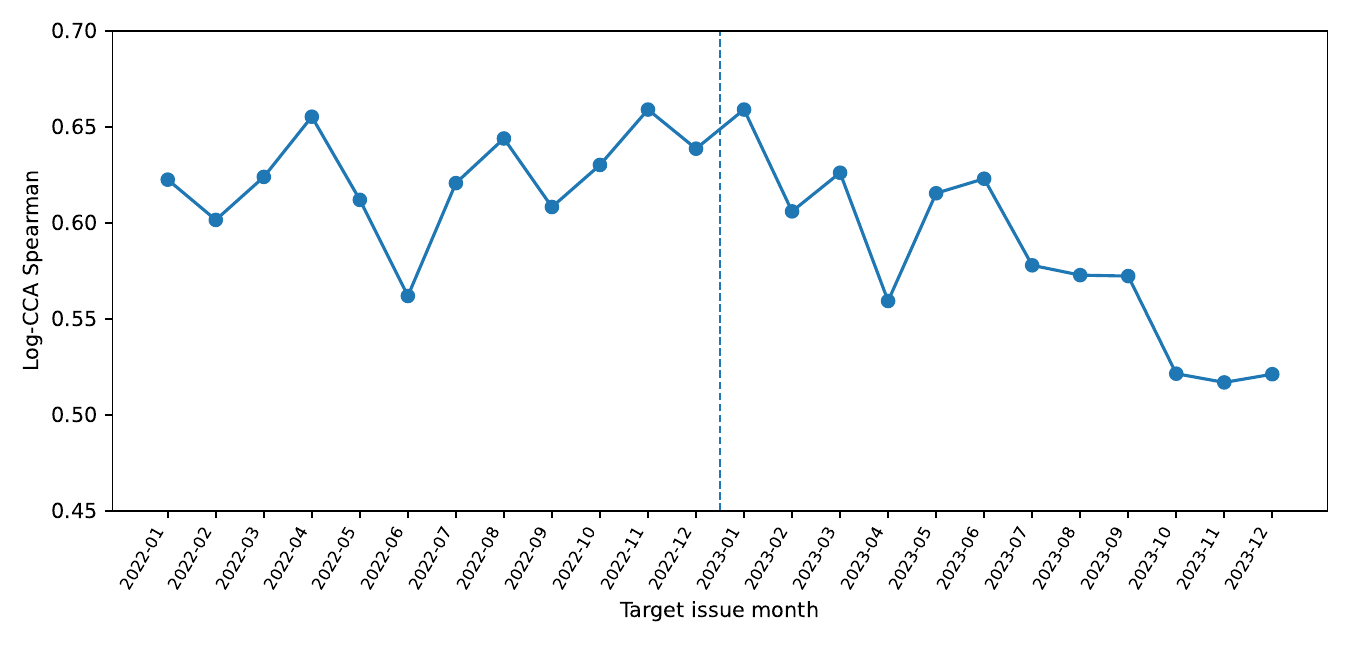}
\caption{Month-ahead frozen log-CCA correlation under the continuous maturity-aware outcome stream.}
\label{fig:monthly}
\end{figure}

The monthly evidence suggests an operational division of labour. Scores can be monitored monthly for drift in rank correlation, positive-outcome prevalence, persistence, and concentration. Full calibration need not be refitted every month if diagnostics remain within tolerance; quarterly recalibration offers more stable data and aligns with procurement planning cycles.

\subsection{Pre-specified week-to-week samples}

Weekly windows are the most demanding tests. The sixteen pre-specified samples have a median Spearman correlation of 0.526 and an interquartile range of 0.489--0.596. The range is wide, from 0.327 to 0.720. Median relation persistence is 88.9\% and relief coverage 96.6\%, but each window contains relatively few positive relations and buyer choice sets.

\begin{figure}[H]
\centering
\includegraphics[width=0.92\textwidth]{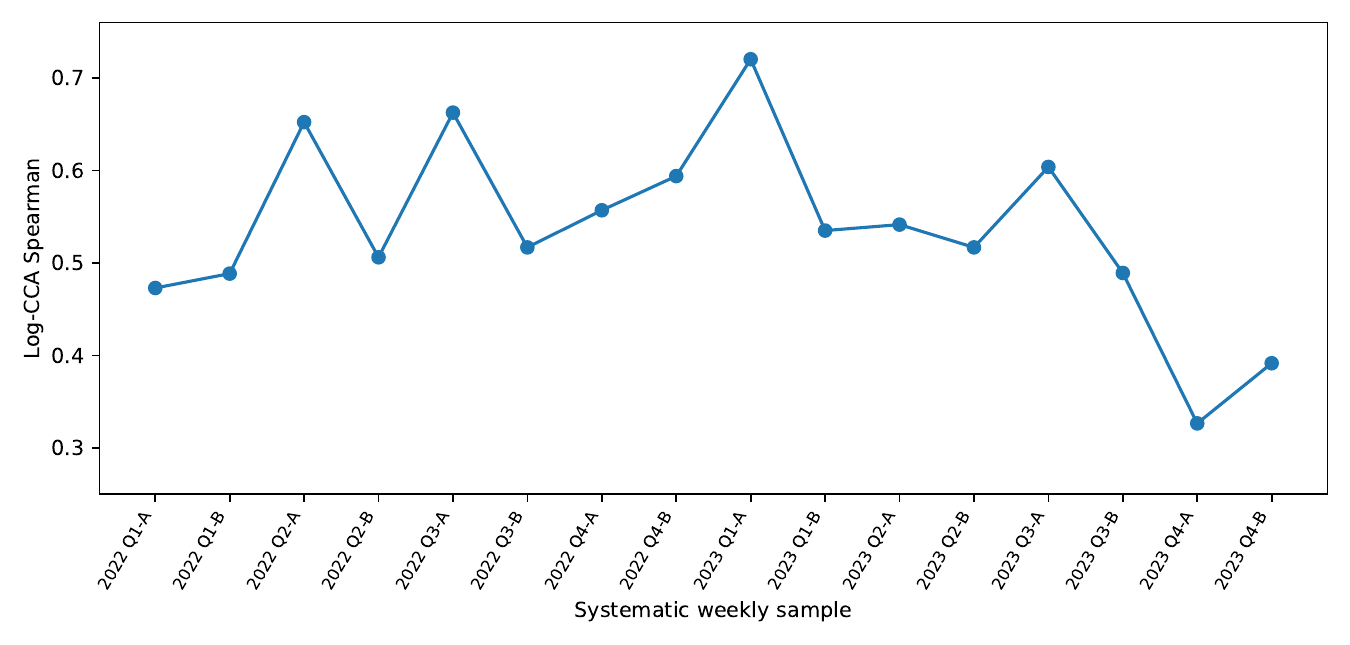}
\caption{Frozen log-CCA Spearman correlation in two pre-specified Monday--Sunday samples per quarter.}
\label{fig:weekly}
\end{figure}

The low values in the two sampled weeks of Q4 2023 coincide with the end of the observed archive and should not be generalized. More broadly, weekly scores are better interpreted as alert and stress diagnostics than as a principal supplier-allocation horizon. A one-week outcome is too sensitive to exact issue timing, short-lived bottlenecks, and whether a large transaction happens to arrive inside the selected seven days.

\subsection{Semester and annual windows}

The complete semester correlations are 0.640 for H1 2022, 0.672 for H2 2022, and 0.626 for H1 2023. H2 2023 has only partial observed follow-up and yields 0.589. The stronger H2 2022 result is consistent with the structurally unusual cycle-saturated 2022 regime, but it is not treated as evidence that the second half of every year is easier to forecast.

\begin{figure}[H]
\centering
\includegraphics[width=0.72\textwidth]{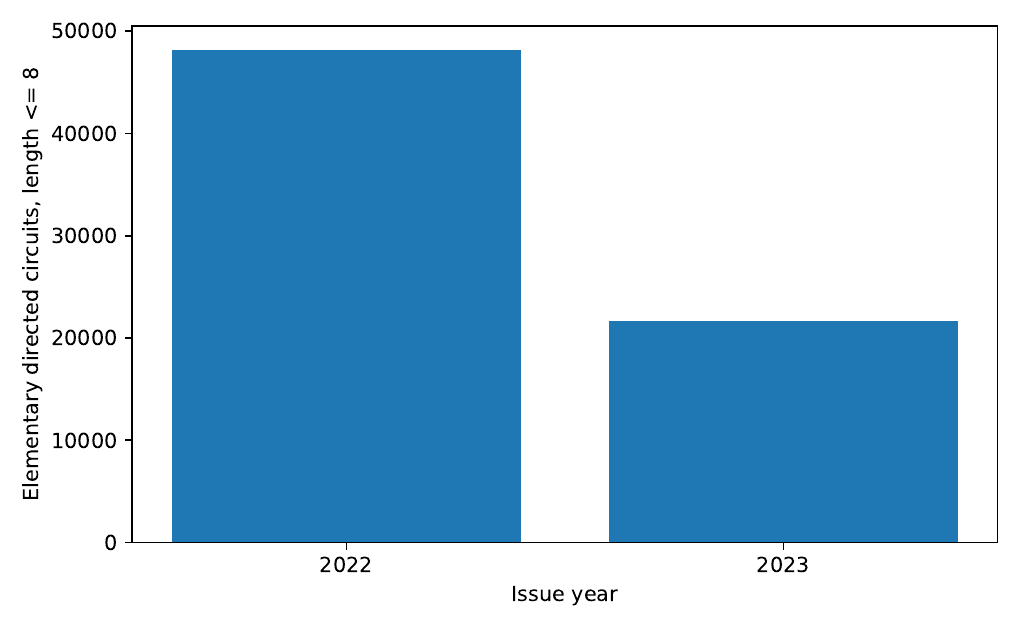}
\caption{Annual elementary directed circuits through length eight. The second half of 2022, including Q4, is interpreted within an unusually cycle-saturated annual regime rather than as a singular forecasting event.}
\label{fig:cycle_context}
\end{figure}

The single complete annual origin, 2022, produces 0.681 Spearman and 0.654 log-Pearson. These values are promising but insufficient to establish annual superiority. Annual forecasting is best treated as a strategic exposure view, while shorter cadences update the operational decision layer.

\subsection{Sequential temporal calibration}

The sequential hurdle model uses only prior completed quarters. The full-control plus log-CCA specification retains positive performance in every test quarter. Spearman correlations range from 0.738 to 0.811; log-$R^2$ ranges from 0.327 to 0.484; and buyer-specific best-relief capture ranges from 84.0\% to 99.5\%.

\begin{table}[H]
\centering
\caption{Sequential full-control plus log-CCA calibration.}
\label{tab:sequential}
\small
\resizebox{\textwidth}{!}{%
\begin{tabular}{lrrrrrr}
\toprule
Target & Training quarters & Spearman & Log-Pearson & Log-$R^2$ & Best capture & Uplift vs mean \\
\midrule
Q2 2022 & 1 & 0.788 & 0.705 & 0.443 & 84.0\% & 216.4\% \\
Q3 2022 & 2 & 0.811 & 0.742 & 0.484 & 99.1\% & 568.6\% \\
Q4 2022 & 3 & 0.781 & 0.700 & 0.414 & 99.5\% & 459.7\% \\
Q1 2023 & 4 & 0.768 & 0.670 & 0.344 & 97.4\% & 285.1\% \\
Q2 2023 & 5 & 0.754 & 0.668 & 0.327 & 96.9\% & 262.9\% \\
Q3 2023 & 6 & 0.772 & 0.678 & 0.384 & 96.6\% & 354.9\% \\
Q4 2023$^{a}$ & 7 & 0.738 & 0.657 & 0.327 & 98.4\% & 422.6\% \\
\bottomrule

\end{tabular}}
\begin{minipage}{0.94\textwidth}\footnotesize
The target quarter is excluded from training. $^{a}$Q4 2023 has partial observed follow-up.
\end{minipage}
\end{table}

H6 is supported. The calibrated model improves magnitude prediction over the raw score but does not eliminate temporal variation. The appropriate deployment architecture is therefore layered: a stable and interpretable score, a periodically refitted calibration model, and monitoring that can trigger earlier review.

\subsection{Overall assessment of hypotheses}

H1 is supported because all eight quarter targets produce a positive prospective association, with seven fully observed quarters forming the primary inference set. H2 is supported by repetition across Q1--Q4 2022 and Q1--Q4 2023 rather than dependence on one favorable period. H3 is supported in the quarter-ahead comparison: Log-CCA has the strongest median combination of rank and log-magnitude alignment among the tested CCA variants, while CPM remains a strong size-neutral structural benchmark. H4 is supported, although buyer-level capture varies by quarter. H5 is supported because Log-CCA predicts thirty-day relief and relief-days, and high-score relations have substantially higher short-term incidence; the weaker conditional waiting-time coefficient appropriately limits exact-date claims. H6 is supported by positive sequential performance and continuing variation in calibration quality. RQ1 reveals a smoothing--coverage--timeliness trade-off rather than a universal rule that one horizon is always best.

\section{Discussion}

\subsection{Prospective validity and temporal transport}

The central empirical result is not the existence of a high contemporaneous correlation. It is that information frozen before a target period retains predictive value for later relief across repeated quarters and multiple horizons. This distinction matters because supplier-selection decisions are made before the future invoice and its compensation opportunities exist. The results therefore move CCA from a retrospective descriptive index toward a prospective decision feature.

The signal is stable enough to recur but not invariant enough to justify a timeless calibration. Quarter-level correlations remain positive throughout 2022--2023, yet their magnitude changes. The sequential models also remain positive while varying in fit. Network participation, invoice concentration, maturity distribution, and residual competition alter the mapping from a structural score to realized euros. The empirical pattern supports a stable construct with a periodically updated calibration layer.

\subsection{Theoretical contribution to supplier selection}

The paper extends supplier-selection theory by treating each qualified supplier as a possible network-financial entry point. Conventional criteria describe the supplier or dyad. CCA adds a relational externality: the contemplated payable can interact with obligations beyond the focal contract. The value of the relation therefore depends on maturity-aware network position and on the capacity to activate that position.

The result is not another generic centrality score. CPM represents expected compensability under specified event channels. CCA then introduces capacity as a complement rather than as a substitute. The quarter results show why that distinction matters. CPM remains strong in structural concentration tests, but Log-CCA is the better balanced operational predictor and buyer-conditioned selector. This division of labour is theoretically useful: structure determines whether relief is possible, whereas capacity determines whether that possibility can be economically material.

\subsection{Relief amount and compensation timing}

The timing extension changes the interpretation of value. Relief is not only a terminal amount; it also has duration. Removing or transforming an obligation soon after issue can reduce the period during which gross settlement exposure occupies working capital, payment capacity, credit limits, and managerial attention. Relief-days make that temporal quantity explicit without equating it automatically with interest savings.

The empirical distinction is equally important. Log-CCA predicts the amount generated within thirty days and the combined amount-duration outcome more strongly than it predicts exact delay conditional on an event. In practical terms, high CCA values identify a favorable short-term opportunity set, but the operation date still depends on when supporting invoices arrive and how competing candidates consume residual capacity. A dashboard should therefore present a probability or expected amount for a defined horizon rather than a promised compensation date.

The apparent speed of the experimental operations also requires institutional interpretation. An algorithmic event can extinguish a cycle claim or transform a path claim on its recorded execution date, but a generated instruction may be paid later. Future field validation must distinguish proposal, consent, legal discharge, instruction issuance, payment initiation, and creditor receipt. The current results establish timely compensability, not universally faster cash payment.

\subsection{Two compensation environments, one measurement question}

The paper deliberately avoids turning the analysis into a contest between cycle and path algorithms. The companion study addresses that policy comparison. Here, the two methods are valuable because they generate relief from different local structures. A measure that correlates only with one mechanically related outcome would have limited construct coverage. Positive results across both channels show that CCA is not tied exclusively to reciprocal closure or to local path reach.

The integrated outcome should likewise be interpreted carefully. It is a multi-environment validation target, not a claim that separate policy totals can be executed jointly. A production platform would require one joint allocator, policy priorities, consent rules, and residual-capacity governance. The timing of the integrated label is consequently used for predictive coverage, not as the chronology of one combined settlement engine.

\subsection{Forecast horizon and decision cadence}

The week-to-year comparison suggests that horizon is part of the estimand, not merely a reporting choice. Weekly windows expose high-frequency fragility. Monthly windows are sufficiently populated for monitoring but remain sensitive to concentration. Quarterly windows balance repeated validation, stable correlations, high relief coverage, timing information, and procurement relevance. Semester windows stabilize the signal further but reduce relation persistence. Annual windows may represent strategic exposure, yet the present archive supplies only one complete annual origin.

Quarterly recalibration is therefore the most defensible default, with monthly monitoring and weekly exception diagnostics. This is an operational recommendation rather than a universal law. Categories with short order cycles, stable counterparties, or denser event streams may support faster updating. Low-frequency categories may require semesters. The methodology makes the trade-off measurable in each deployment.

\subsection{The 2022 structural context}

The only special interpretation attached to Q4 2022 is that it falls within an unusually reciprocal and cycle-saturated second half of 2022. Annual reciprocity is 68.88\%, the largest strongly connected component contains 35.11\% of firms, and the graph has 48,109 elementary circuits through length eight. The second half of 2022 contains roughly twice as many issue-window circuits as the second half of 2023 in the target-topology audit.

That context is useful because CCA remains predictive even when path and cycle opportunities overlap strongly. It does not justify treating Q4 2022 as a uniquely favorable or representative quarter. The full Q1--Q4 sequences in both years are the proper evidence.

\subsection{Production-economic and supply-chain-finance implications}

Supplier selection determines where purchasing demand becomes a liability. When the placement changes eventual payable-mass reduction, the decision affects gross settlement exposure before financing instruments are applied. CCA therefore adds an endogenous-relief stage to the usual working-capital sequence: estimate compensability and its timing, execute governed reduction where accepted, and finance the residual.

The measure is not itself liquidity and does not prove a lower cost of capital. PMR reduces gross claims that must be settled, while relief-days approximate how long this reduction precedes maturity. Net funding benefit still depends on acceptance, legal discharge, fees, payment acceleration, counterparty exposure, actual cash balances, and the availability and price of external finance. A causal field study must observe these downstream outcomes.

\subsection{Competitive advantage and organizational capability}

The formula is transparent and therefore unlikely to be a sustainable advantage by itself. The potentially scarce resource is the capability system around it: broad and accurate invoice visibility, identity resolution, maturity and residual data, participation density, legal and policy eligibility, procurement-category context, and feedback from accepted and rejected operations.

A firm with these complements can use CCA in three ways. First, it can differentiate otherwise acceptable suppliers by expected future relief and timely-compensation opportunity. Second, it can coordinate procurement and treasury around the same network signal. Third, repeated decisions and outcomes can generate proprietary learning about which relations remain predictive under changing regimes. Competitors lacking comparable data, partner participation, or governance may copy the equation without reproducing the capability.

\subsection{Managerial implementation architecture}

A practical implementation should contain four layers. The first is a deterministic data layer that verifies identifiers, outstanding amounts, issue dates, due dates, and residual continuity. The second is a structural scoring layer that calculates CPM and Log-CCA from lagged information. The third is a calibration layer trained only on completed periods and capable of producing horizon-specific relief and timing estimates. The fourth is a governance layer that applies qualification, consent, exposure, acceleration, legal, audit, and fairness constraints.

Decision dashboards should report the score together with its horizon, origin date, expected relief, probability of short-term compensation, relief-days, persistent-relief coverage, recent calibration performance, and drift status. A single timeless CCA number would obscure the temporal and institutional conditions under which it was estimated.

\section{Future applications and scientific and societal implications}

\subsection{Scientific impact: from static centrality to dated economic opportunity}

The study contributes a temporal-network interpretation of supplier choice. Static centrality asks whether a node is prominent in an aggregated graph. CCA asks a more operational question: whether a contemplated relation connects the buyer to obligations that are simultaneously active, sufficiently funded by residual capacity, and likely to support measurable relief within a defined future horizon. The distinction is scientifically relevant because many networks are stateful: using one opportunity changes the capacity available to every later opportunity.

This perspective can inform research beyond invoice clearing. Temporal graph science often studies reachability or diffusion; the present framework adds an amount-valued, maturity-constrained economic event. Financial-network research often studies default or contagion; CCA studies a pre-default transformation that may reduce gross settlement requirements. Operations research often optimizes a known network; frozen-origin validation asks whether the current network contains information about a future operational outcome. These intersections create a research program in which topology, timing, capacity, and decision value are estimated jointly rather than sequentially.

The timing contribution is especially extensible. Relief-days is one instance of a broader class of duration-weighted network outcomes. Similar measures could weight avoided inventory exposure by days, energy balancing by hours, transport consolidation by lead-time reduction, or emissions abatement by the duration for which a high-carbon activity is displaced. The scientific challenge is to define the conserved quantity, the admissible event, the time origin, and the counterfactual duration without confusing predictive association with causal welfare.

\subsection{Methodological transfer to other stateful systems}

The frozen-origin architecture separates four objects: the information available when a score is calculated, the physical state on which operations act, the future cohort being evaluated, and the horizon over which outcomes are observed. That separation is useful wherever interventions consume shared capacity and overlapping evaluation windows would otherwise reuse the same resource. Examples include credit-limit allocation, matching markets, hospital capacity, transport slots, renewable-energy flexibility, and inventory pooling.

The paper also separates three validation questions that are frequently collapsed into one. Rank validity asks whether high scores locate better opportunities. Magnitude calibration asks whether the score can be converted into a defensible expected amount. Decision validation asks whether following the score improves the actual constrained choice. Timing adds a fourth question: whether the score identifies outcomes soon enough to matter. A measure may perform well on one dimension and poorly on another; reporting all four prevents a favorable global correlation from masking weak buyer-level or temporal performance.

Future methodological work can use survival analysis, recurrent-event models, temporal point processes, and dynamic graph neural networks to model compensation timing. Such models should be benchmarked against transparent CPM and Log-CCA rather than replacing them by default. Their added complexity is justified only if they improve out-of-time calibration, buyer-level regret, or institutional decision quality while preserving explainability and avoiding leakage.

\subsection{Procurement, treasury, and supplier-portfolio applications}

The immediate application is a network-aware extension of supplier qualification and order allocation. CCA should enter only after suppliers satisfy category, price, quality, delivery, sustainability, compliance, and resilience requirements. Within that qualified set, procurement can use the score to identify which placement is more likely to generate payable relief and to do so within a decision-relevant horizon. Treasury can use the same signal to estimate which obligations may be reduced before external funding is arranged.

The timing outputs support several operating choices. A high probability of thirty-day compensation may justify waiting before drawing short-term finance, subject to liquidity buffers and legal certainty. High relief-days can identify relations whose early settlement transformation would free gross exposure for longer. Low timing confidence can trigger conservative funding or manual review even when eventual relief is plausible. These are decision-support uses, not automatic instructions; they require risk limits and a clear distinction between claim transformation and cash receipt.

At portfolio level, the measure can support dual sourcing and spend allocation. Rather than replacing an operationally superior supplier, a buyer could allocate the financially discretionary share of an order toward a higher-CCA alternative. Scenario analysis can impose minimum shares, concentration limits, resilience constraints, and switching costs. The resulting optimization would test the marginal working-capital value of network placement while retaining conventional procurement objectives.

\subsection{Circular supply chains and industrial symbiosis}

Circular supply-chain research primarily concerns the narrowing, slowing, and closing of material and energy loops \cite{farooque2019}. Financial circularity is different: it concerns the ability of dated obligations to be reduced through reciprocal or progressive network structure. The two forms of circularity need not coincide. A supplier can perform strongly on material recovery yet occupy a weak compensability position, or support financial relief without contributing to material circularity.

Their combination nevertheless creates a promising application. Circular business models often require up-front investment, reverse-logistics coordination, refurbishment capacity, and new interfirm relationships. Working-capital pressure can slow adoption, particularly for smaller suppliers. A joint supplier dashboard could therefore report material-circularity indicators alongside CPM, Log-CCA, timely-relief probability, and relief-days. The purpose would not be to redefine environmental circularity as finance, but to identify supplier configurations that are both environmentally circular and financially easier to sustain.

Industrial symbiosis offers a more specific case. Firms exchange by-products, heat, water, logistics capacity, or recovered materials, creating networks in which one firm's residual becomes another's input. These exchanges can generate reciprocal commercial claims as well as physical loops. CCA could help test whether a proposed symbiosis cluster also creates compensable financial structure, thereby reducing one barrier to participation. A valid study would need to measure whether predicted financial relief causally increases adoption, continuity, or environmental performance rather than assuming that liquidity automatically produces circular outcomes.

Circularity benchmarking remains fragmented and indicator-rich \cite{dekoning2024}. CCA contributes a narrowly defined financial complement: expected gross-obligation reduction and its timing. It should not be aggregated into an opaque universal circularity index. Keeping the material, environmental, operational, and financial dimensions visible would allow researchers and managers to examine trade-offs rather than hide them inside one score.

\subsection{Platform ecosystems, agentic coordination, and privacy-preserving clearing}

Digital B2B platforms can use CCA as a routing and participation signal. A platform can identify relations with high latent compensability, invite missing counterparties, prioritize data-quality improvements, or present governed clearing opportunities. Because the economically relevant core is more persistent than the full relational perimeter, a platform may obtain meaningful value before universal participation is achieved.

An agentic implementation could represent firms with constrained software agents that search for admissible opportunities, negotiate discretionary terms, and request deterministic verification. The score can guide discovery, while authoritative capacity, residual mutation, and conservation remain outside the language model. Such a system would need signed mandates, short-lived reservations, atomic commit, finality rules, and replayable evidence. CCA supplies a prioritization layer; it does not by itself solve consent, strategic withholding, privacy, or legal discharge.

Privacy-preserving computation is particularly relevant because complete invoice-network visibility is commercially sensitive. Secure multiparty computation, trusted execution environments, zero-knowledge commitments, federated feature calculation, or local disclosure of only the necessary capacity certificate could permit partial deployment without exposing the full graph. Future experiments should report not only predictive performance but also disclosure volume, communication cost, stale reservations, accepted relief, time to finality, and fairness among participants.

\subsection{Public-sector, SME, cooperative, and community applications}

The same logic may apply to public-sector arrears, health-service purchasing, municipal suppliers, agricultural cooperatives, telecommunications settlements, franchise systems, intercompany treasury, and community-energy networks. In each case, organizations can be creditors and debtors at the same time, and late settlement can propagate operational pressure. A network-aware measure could help identify claims that are likely to be reduced collectively before scarce external finance is used.

The potential SME impact is significant but must be tested carefully. Small suppliers generally face higher financing constraints and less bargaining power. Earlier compensation could reduce the duration of gross exposure, yet a capacity-adjusted score may also favor established high-volume firms. A responsible implementation should therefore separate predictive ranking from procurement eligibility, provide cold-start treatment for new suppliers, impose concentration and diversity constraints, and audit whether recommendations shift value away from smaller or peripheral firms.

Cooperative and public applications raise governance questions beyond private efficiency. Who may see the network? Who defines eligibility? Can a participant refuse redirection without losing future business? How are disputes and errors reversed? Are savings distributed proportionately, or captured by the platform and the largest firms? A societally valuable system requires transparent rules, independent oversight, accessible appeals, and explicit allocation of the benefits and risks of earlier compensation.

\subsection{Resilience, early warning, and macro-network applications}

CCA may also serve as an early-warning feature. A declining score, falling persistent-relief coverage, or lengthening predicted compensation term can indicate that a supplier network is losing its ability to absorb gross obligations endogenously. Combined with payment-delay, concentration, and credit indicators, such changes could help treasury teams identify emerging working-capital pressure before it appears in default statistics.

At a system level, aggregates of compensability and relief-days could complement conventional measures of trade-credit exposure. They may help distinguish a network with large gross obligations but strong internal reduction capacity from one in which similar gross volume must be funded externally. This is relevant to resilience research, which emphasizes that network structure can transmit or absorb disruption \cite{ivanov2020}. It remains an empirical question whether higher compensability reduces distress propagation, merely rearranges exposures, or creates new concentration and dependency risks.

Regulators and infrastructure operators could use anonymized aggregates to assess how participation rules, payment standards, or interoperability affect endogenous relief. Such use should remain diagnostic. The present data are not demonstrated to be nationally representative, and PMR is not welfare. A macro-policy claim requires broader jurisdictions, sectoral coverage, legal analysis, and evidence on actual cash, financing, default, and distributional consequences.

\subsection{Societal safeguards and responsible deployment}

The social value of CCA depends on who benefits and who bears the cost. A platform optimized only for aggregate relief may concentrate transactions around already central firms, disclose sensitive commercial relationships, accelerate some payers, or exclude new suppliers that lack history. Predictive accuracy is therefore necessary but not sufficient. Deployment should include fairness constraints, counterparty exposure limits, competition safeguards, and reporting by supplier size, sector, geography, and network position.

Explainability is unusually feasible because the score has separable components. A decision can state whether it is driven by structural opportunity, bilateral capacity, short-horizon likelihood, or recent calibration. That decomposition should be retained even if more complex models are added. It gives procurement teams a basis for challenge and allows suppliers to understand why a relation receives a particular ranking.

Data governance must include purpose limitation, least-privilege access, retention rules, secure entity resolution, and protection against re-identification from topology. Consent and legal finality must be explicit. A high-scoring opportunity is not authority to alter a claim. The scientific agenda should therefore evaluate accepted relief, not only algorithmic capacity, and treat refusals, disputes, reversals, and failed payments as first-class outcomes.

\subsection{Application map and future research program}

Table~\ref{tab:application_map} summarizes the broader application space. The common principle is to use CCA as a transparent predictive feature inside a governed decision system, not as an autonomous objective that overrides operational, environmental, legal, or social constraints.

\begin{table}[H]
\centering
\caption{Illustrative future applications of CCA and the evidence required for responsible use.}
\label{tab:application_map}
\footnotesize
\begin{tabularx}{\textwidth}{L{2.4cm}YYL{3.0cm}}
\toprule
Domain & Decision supported & Candidate outcome & Essential safeguard or research test \\
\midrule
Procurement and treasury & Allocate qualified spend and time residual funding & Accepted relief, relief-days, borrowing duration, buyer regret & Category feasibility, liquidity buffers, causal field test \\
Circular supply chains & Combine material-circularity and financial-compensability views & Adoption, continuity, working-capital exposure, environmental outcomes & Keep environmental and financial dimensions separate; test causality \\
Industrial symbiosis & Design reciprocal exchange clusters & Cluster participation, claim reduction, project survival & Contract enforceability, benefit sharing, partner concentration \\
B2B platforms & Prioritize onboarding, data quality, and opportunity discovery & Accepted PMR, participation, time to finality & Privacy, consent, transparent fees, interoperability \\
Public and cooperative networks & Reduce arrears and coordinate claims under common rules & Payment delay, service continuity, SME liquidity & Due process, equal access, public oversight, distributional audit \\
Resilience and early warning & Monitor deterioration in endogenous relief capacity & Funding pressure, delay propagation, distress incidence & Multi-market validation and separation from causal systemic-risk claims \\
Agentic coordination & Support local proposal and negotiation among constrained agents & Accepted relief per message, latency, failure, disclosure & Deterministic verification, signed mandates, atomic commit, replay \\
\bottomrule
\end{tabularx}
\end{table}

The next scientific phase should connect four levels of evidence. First, extend out-of-time prediction over additional complete years and jurisdictions. Second, record proposal, consent, legal discharge, payment initiation, and creditor receipt so that claim-transformation timing can be separated from cash timing. Third, run category-qualified procurement interventions that estimate causal effects on working capital, price, quality, resilience, and supplier outcomes. Fourth, test whether repeated CCA-guided choices alter network topology, creating either beneficial financial circularity or undesirable concentration. This sequence would move the research from predictive validation to institutional and societal evaluation.

\section{Limitations and research agenda}

The source archive is one institutional environment. Coverage changes across years, and firms, relations, and invoices are not independent observations. Buyer-cluster bootstrap intervals are stability diagnostics rather than population confidence intervals. Q4 2023 and all longer windows requiring later 2024 observations are right-censored because the available source ends on 8 February 2024.

The two outcome environments are deterministic algorithmic labels. They establish whether the score anticipates structurally feasible relief under specified policies, not whether every proposed operation would be accepted or legally final in practice. The integrated label sums independently generated opportunities and is not a feasible joint policy. A future study should estimate CCA against a governed mixed allocator and compare predicted with accepted relief.

The timing analysis observes the operation that receives source-fragment PMR attribution. It does not observe every institutional stage between opportunity discovery and cash receipt. Cycle settlement, claim discharge, generated instruction, payment initiation, payment finality, and creditor access to funds can occur on different dates. The reported issue-to-event delay and days before maturity should therefore be interpreted as timing of compensability or claim transformation, not as a direct estimate of days-sales-outstanding reduction.

Relief-days cap maturity lead at 365 days to limit the influence of a very small number of extreme due dates. The uncapped aggregate differs by only 0.37\% in the path environment and 0.20\% in the cycle environment, but alternative weighting functions should be tested. Future work should model time to accepted relief using survival or recurrent-event methods and should evaluate calibration at 7-, 30-, 60-, and 90-day horizons.

The path implementation is independently coded and does not reproduce every companion value exactly. The cycle stream reconciles to less than 0.00005 percentage points, while path differences reach 0.371 percentage points because deterministic local ordering is not fully unique. Future releases should define one canonical tie-breaking specification and source-level replay format shared by both papers.

The capacity measures use historical observed invoices. They do not incorporate contractually committed future volume, supplier willingness, or category substitutability. Buyer-specific choice sets are observed relations rather than randomized procurement alternatives, so the paper does not establish causal benefit from following the recommendation. New and low-volume suppliers also face a cold-start problem that requires either priors, category-level features, or protected exploration.

The horizon comparison has unequal numbers of complete origins: sixteen weeks, twenty-three months, seven quarters, three semesters, and one year. The apparent rise in correlation with horizon may reflect both smoothing and origin composition. Longer data through at least 2025 would provide multiple complete annual and semester origins and allow a hierarchical model of horizon effects.

Economic causality remains untested. Payable-mass reduction and relief-days are not identical to cash saving, interest saving, profitability, resilience, or social welfare. The conversion depends on actual liquidity, funding costs, fees, legal discharge, acceleration imposed on payers, counterparty exposure, default, tax treatment, and supplier response. A phased or randomized field implementation should measure these outcomes alongside procurement price, quality, delivery, supplier acceptance, concentration, fairness, and regret.

Future work should therefore pursue eight priorities: additional complete years and jurisdictions; canonical cross-paper execution and replay; direct observation of payment finality; survival modelling of compensation timing; category-qualified supplier-choice experiments; adoption-adjusted relief net of fees, acceleration, and risk; integration with environmental circularity and industrial-symbiosis studies; and distributional audits covering SMEs, new suppliers, peripheral firms, and public-interest deployments.

\section{Conclusion}

This paper asks whether information available before a purchasing period can identify supplier relations that are more likely to generate future payable relief, and whether that relief is likely to arise soon enough to matter for working-capital exposure. The answer is affirmative but conditional. A dated invoice network contains a repeatable prospective signal, yet the strength and calibration of that signal depend on horizon, network regime, relation persistence, commercial capacity, and the institutional meaning of a compensation event.

The first contribution is conceptual. Supplier selection is normally evaluated through attributes of the supplier or contract. CCA adds a network-financial consequence: a new payable enters a temporal system of receivables and payables, and its placement changes the opportunities through which gross obligations may be reduced before residual settlement or external funding. This consequence is relational rather than purely dyadic. It depends on maturity-compatible topology, bottleneck capacity, and the supplier's ability to activate that structure at meaningful scale.

The second contribution is measurement. Expected relief is the underlying monetary estimand. CPM is its bounded, size-neutral structural representation. Raw CCA exposes the risk of multiplying that signal by an unrestricted capacity ratio in a heavy-tailed commercial network. Concave transformations control that leverage, with Log-CCA providing the strongest operational compromise in the repeated quarter-ahead tests. CPM remains valuable as the transparent structural kernel; Log-CCA adds an interpretable commercial activation layer. Neither should be treated as an invariant euro forecast without calibration.

The third contribution is genuinely prospective evidence. Across Q1--Q4 2022 and Q1--Q3 2023, the seven fully observed quarters, frozen Log-CCA has Spearman correlations between 0.592 and 0.663 with future integrated relief and a median of 0.638. The association is positive in the final right-censored quarter as well, although it is not included in the primary median. Persistent relations carry a median 93.5\% of future relief, indicating that most economically material compensability is concentrated in a relatively stable relational core even while the network perimeter changes.

The fourth contribution is decision relevance. Global correlation does not by itself establish that a procurement rule chooses well inside each buyer's feasible set. The buyer-conditioned experiment shows that selecting the highest frozen Log-CCA relation captures a median 92.5\% of buyer-specific best relief and produces a substantial uplift over the mean candidate. Decision quality varies across quarters, so the score is not infallible. Nevertheless, the result demonstrates that the network signal survives the more demanding comparison among a buyer's own alternatives.

The fifth contribution is temporal. Across the two independent validation environments, approximately 89--91\% of attributed relief is generated within seven days of invoice issue and 94--96\% within thirty days. The PMR-weighted mean event occurs roughly four days after issue and about fifteen to sixteen days before contractual maturity. Log-CCA correlates 0.564 with integrated thirty-day relief and 0.566 with integrated relief-days across the complete quarters. Its highest quartile has a 96.5\% median incidence of positive thirty-day compensation, compared with 57.6\% in the lowest quartile. Selecting the highest-Log-CCA relation captures 94.7\% of buyer-specific best thirty-day relief at the median.

These timing findings are important precisely because they are bounded. CCA is much better at identifying whether economically material relief is likely to occur within a useful short horizon than at predicting the exact day conditional on an event. The median conditional waiting-time correlation is only 0.209 for the integrated target. Moreover, an experimental compensation date is not always a creditor cash-receipt date. The valid claim is that CCA ranks timely compensation or claim-transformation opportunity and the duration of gross exposure avoided. Claims about days-sales-outstanding, financing cost, or cash arrival require direct operational data.

The sixth contribution is horizon discipline. Predictive association is positive from week to year, but each horizon serves a different decision. Weekly estimates are volatile and best used for exception diagnostics. Monthly estimates support drift monitoring. Quarterly estimates offer the strongest balance of repeated evidence, signal, coverage, timing, and procurement cadence. Semester estimates support portfolio review. The annual result is strategically informative but rests on one fully observed origin. The evidence therefore supports quarterly recalibration, monthly monitoring, and weekly diagnostics rather than leaving one model unchanged for a year.

The seventh contribution is construct coverage. Cycle-restricted and path-enabled compensation generate relief from different local structures. This paper does not compare their superiority; it uses them as independent tests of whether CCA remains informative under reciprocal closure and local path opportunity. Positive results in both environments show that the measure is not tied to one settlement primitive. Their integrated label broadens validation coverage but is not a jointly executable total.

The broader production-economic implication is that endogenous relief can be considered before external finance. A firm can first estimate which qualified supplier placements are likely to reduce gross obligations and when that reduction may occur, then apply governed compensation where accepted, and finance the residual. This sequence does not eliminate the need for credit. It adds a network-aware decision stage that can reduce unnecessary gross settlement pressure and improve coordination between procurement and treasury.

The circularity implication is similarly specific. Financial circularity is not a substitute for material circularity, environmental performance, or circular-economy design. It is a complementary property describing how effectively a dated obligation network can reduce gross claims. Combining environmental and financial indicators may help identify circular supplier configurations that are both resource-efficient and financially sustainable, but that possibility must be tested rather than assumed.

The societal potential extends to SME networks, cooperatives, public-sector arrears, industrial symbiosis, platform ecosystems, and community infrastructure. Earlier and more predictable compensation could reduce the duration of working-capital pressure and improve continuity. Yet the same measure could favor established high-volume firms, increase concentration, or expose sensitive topology if deployed without safeguards. Responsible use therefore requires qualification rules, cold-start treatment, fairness and concentration constraints, privacy-preserving data access, explicit consent, legal finality, transparent fees, and independent oversight.

The strategic advantage does not lie in keeping the equation secret. It can arise from a difficult-to-replicate capability system: broad invoice visibility, accurate maturity and residual data, interoperable identities, sufficient participation, governed execution, procurement-category knowledge, treasury integration, and recurrent learning from accepted and rejected outcomes. A competitor may copy the formula without reproducing the data, relationships, controls, and organizational routines that make the estimate useful.

The study establishes predictive validity and supplier-prioritization value, not causal financial or social impact. The decisive next step is a prospective field intervention with qualified alternatives. It should record proposed and accepted compensation, legal discharge, instruction issuance, payment finality, creditor receipt, funding use and duration, fees, acceleration imposed on payers, procurement performance, supplier response, concentration, and fairness. Repeated studies across jurisdictions and sectors should then test whether CCA-guided decisions alter network structure, working-capital resilience, circular-business adoption, or distributional outcomes.

The final conclusion is therefore both stronger and narrower than a generic claim that network circularity is beneficial. Financial compensability can be estimated before the future period; its amount and short-term incidence are measurably related to supplier-network placement; and the signal can improve buyer-specific prioritization when capacity is regularized and calibration is updated. Used as one governed criterion alongside price, quality, delivery, resilience, sustainability, compliance, and risk, CCA provides a practical way to make the financial consequence of circular network placement visible, testable, and actionable.

\section*{Data and code availability}

The pseudonymized experimental database is available as: Esteva de la Rosa, Peplluis (2026), \emph{Atomic Common-Day Invoice Clearing: Pseudonymized Invoice Records and Reproducibility Data, 2012--2023}, Mendeley Data, Version 1, doi: \href{https://doi.org/10.17632/28rbmvwsm9.1}{10.17632/28rbmvwsm9.1} \cite{esteva2026data}. The accompanying GitHub-ready repository \href{https://github.com/peplluis7/CCA-compensability} {github.com/peplluis7/CCA-compensability} contains the source, atomic-data preparation scripts, continuous-state outcome engines, frozen-history builder, pre-specified origin definitions, CCA variants, relief and compensation-timing analysis, quarterly bootstrap analysis, sequential calibration, aggregate tables, figures, tests, and verification workflows. Raw identifiable records and reversible identity mappings are not distributed.

\section*{Declaration of generative AI and AI-assisted technologies}

During preparation of this work, the authors used an AI-assisted language and coding system to support editing, code review, artifact generation, and consistency checks. The authors reviewed the resulting text, formulas, code, and empirical outputs and remain responsible for the content of the work.

\appendix

\section{Complete quarterly formula comparison}
\label{app:variant_detail}

\begin{longtable}{llrrr}
\caption{All formula variants against future integrated relief by quarter.}\label{tab:quarter_variants_full}\\
\toprule
Quarter & Score & Spearman & Log-Pearson & Top-decile capture \\
\midrule
\endfirsthead
\toprule
Quarter & Score & Spearman & Log-Pearson & Top-decile capture \\
\midrule
\endhead
Q1 2022 & CPM & 0.563 & 0.552 & 68.5\% \\
 & Raw CCA & 0.574 & 0.585 & 77.4\% \\
 & Sqrt CCA & 0.594 & 0.604 & 77.4\% \\
 & Capped CCA & 0.573 & 0.584 & 77.3\% \\
 & Log-CCA & 0.617 & 0.596 & 78.4\% \\
 & LOO CCA & 0.408 & 0.420 & 66.8\% \\
\addlinespace
Q2 2022 & CPM & 0.586 & 0.585 & 79.6\% \\
 & Raw CCA & 0.606 & 0.611 & 61.5\% \\
 & Sqrt CCA & 0.624 & 0.635 & 61.5\% \\
 & Capped CCA & 0.606 & 0.607 & 61.4\% \\
 & Log-CCA & 0.645 & 0.632 & 64.5\% \\
 & LOO CCA & 0.419 & 0.425 & 41.9\% \\
\addlinespace
Q3 2022 & CPM & 0.601 & 0.621 & 94.2\% \\
 & Raw CCA & 0.583 & 0.592 & 82.0\% \\
 & Sqrt CCA & 0.610 & 0.633 & 84.2\% \\
 & Capped CCA & 0.586 & 0.593 & 82.1\% \\
 & Log-CCA & 0.638 & 0.648 & 84.4\% \\
 & LOO CCA & 0.385 & 0.387 & 78.4\% \\
\addlinespace
Q4 2022 & CPM & 0.642 & 0.627 & 95.9\% \\
 & Raw CCA & 0.614 & 0.613 & 77.9\% \\
 & Sqrt CCA & 0.640 & 0.648 & 81.3\% \\
 & Capped CCA & 0.619 & 0.615 & 77.9\% \\
 & Log-CCA & 0.663 & 0.656 & 82.6\% \\
 & LOO CCA & 0.403 & 0.414 & 77.9\% \\
\addlinespace
Q1 2023 & CPM & 0.614 & 0.590 & 91.8\% \\
 & Raw CCA & 0.581 & 0.593 & 61.1\% \\
 & Sqrt CCA & 0.612 & 0.628 & 61.1\% \\
 & Capped CCA & 0.590 & 0.597 & 61.1\% \\
 & Log-CCA & 0.647 & 0.632 & 63.2\% \\
 & LOO CCA & 0.373 & 0.394 & 37.0\% \\
\addlinespace
Q2 2023 & CPM & 0.611 & 0.586 & 86.4\% \\
 & Raw CCA & 0.468 & 0.496 & 43.3\% \\
 & Sqrt CCA & 0.520 & 0.561 & 43.3\% \\
 & Capped CCA & 0.491 & 0.518 & 42.6\% \\
 & Log-CCA & 0.597 & 0.602 & 53.5\% \\
 & LOO CCA & 0.293 & 0.308 & 30.0\% \\
\addlinespace
Q3 2023 & CPM & 0.589 & 0.582 & 91.4\% \\
 & Raw CCA & 0.467 & 0.517 & 60.1\% \\
 & Sqrt CCA & 0.504 & 0.573 & 60.1\% \\
 & Capped CCA & 0.467 & 0.516 & 60.1\% \\
 & Log-CCA & 0.592 & 0.605 & 60.0\% \\
 & LOO CCA & 0.302 & 0.340 & 49.4\% \\
\addlinespace
Q4 2023 & CPM & 0.515 & 0.500 & 90.8\% \\
 & Raw CCA & 0.408 & 0.390 & 58.1\% \\
 & Sqrt CCA & 0.443 & 0.446 & 58.1\% \\
 & Capped CCA & 0.408 & 0.389 & 58.1\% \\
 & Log-CCA & 0.529 & 0.491 & 58.1\% \\
 & LOO CCA & 0.289 & 0.259 & 57.3\% \\
\bottomrule

\end{longtable}

\section{Monthly detail}

\begin{longtable}{lrrrr}
\caption{Month-ahead Log-CCA results under continuous-state outcome attribution.}\label{tab:months_full}\\
\toprule
Month & Persistent relations & Spearman & Log-Pearson & Best relief capture \\
\midrule
\endfirsthead
\toprule
Month & Persistent relations & Spearman & Log-Pearson & Best relief capture \\
\midrule
\endhead
2022-01 & 1244 & 0.622 & 0.558 & 83.1\% \\
2022-02 & 1259 & 0.601 & 0.566 & 88.4\% \\
2022-03 & 1389 & 0.624 & 0.577 & 57.0\% \\
2022-04 & 1190 & 0.655 & 0.631 & 67.0\% \\
2022-05 & 1059 & 0.612 & 0.543 & 63.5\% \\
2022-06 & 708 & 0.562 & 0.418 & 68.7\% \\
2022-07 & 1367 & 0.621 & 0.573 & 84.9\% \\
2022-08 & 1603 & 0.644 & 0.649 & 87.1\% \\
2022-09 & 1612 & 0.608 & 0.595 & 85.6\% \\
2022-10 & 1571 & 0.630 & 0.624 & 80.3\% \\
2022-11 & 1610 & 0.659 & 0.643 & 84.7\% \\
2022-12 & 1730 & 0.639 & 0.600 & 83.8\% \\
2023-01 & 1358 & 0.659 & 0.636 & 82.2\% \\
2023-02 & 1357 & 0.606 & 0.615 & 58.4\% \\
2023-03 & 1589 & 0.626 & 0.575 & 71.0\% \\
2023-04 & 1571 & 0.559 & 0.535 & 70.1\% \\
2023-05 & 1613 & 0.615 & 0.596 & 41.3\% \\
2023-06 & 1452 & 0.623 & 0.626 & 32.6\% \\
2023-07 & 1453 & 0.578 & 0.583 & 56.1\% \\
2023-08 & 1416 & 0.573 & 0.556 & 53.8\% \\
2023-09 & 1317 & 0.572 & 0.531 & 65.9\% \\
2023-10 & 1246 & 0.521 & 0.464 & 53.5\% \\
2023-11 & 964 & 0.517 & 0.496 & 78.9\% \\
2023-12$^{a}$ & 847 & 0.521 & 0.529 & 65.7\% \\
\bottomrule

\end{longtable}

\section{Pre-specified weekly samples}

\begin{longtable}{L{0.24\textwidth}rrrrr}
\caption{Two pre-specified weekly samples per quarter.}\label{tab:weeks_full}\\
\toprule
Sample and start date & Persistent & Spearman & Log-$r$ & Best capture & Buyers \\
\midrule
\endfirsthead
\toprule
Sample and start date & Persistent & Spearman & Log-$r$ & Best capture & Buyers \\
\midrule
\endhead
2022 Q1-A\newline (2022-01-03) & 328 & 0.473 & 0.392 & 53.1\% & 32 \\
2022 Q1-B\newline (2022-02-14) & 511 & 0.489 & 0.436 & 95.0\% & 51 \\
2022 Q2-A\newline (2022-04-04) & 394 & 0.652 & 0.646 & 89.5\% & 32 \\
2022 Q2-B\newline (2022-05-16) & 371 & 0.506 & 0.424 & 77.0\% & 34 \\
2022 Q3-A\newline (2022-07-04) & 467 & 0.663 & 0.644 & 66.5\% & 36 \\
2022 Q3-B\newline (2022-08-15) & 452 & 0.517 & 0.497 & 92.0\% & 35 \\
2022 Q4-A\newline (2022-10-03) & 501 & 0.557 & 0.520 & 91.0\% & 41 \\
2022 Q4-B\newline (2022-11-14) & 564 & 0.594 & 0.532 & 67.8\% & 41 \\
2023 Q1-A\newline (2023-01-02) & 289 & 0.720 & 0.712 & 97.6\% & 22 \\
2023 Q1-B\newline (2023-02-13) & 412 & 0.535 & 0.456 & 67.8\% & 32 \\
2023 Q2-A\newline (2023-04-03) & 499 & 0.542 & 0.536 & 96.5\% & 41 \\
2023 Q2-B\newline (2023-05-15) & 465 & 0.517 & 0.471 & 93.3\% & 35 \\
2023 Q3-A\newline (2023-07-03) & 500 & 0.604 & 0.661 & 41.4\% & 34 \\
2023 Q3-B\newline (2023-08-14) & 344 & 0.489 & 0.394 & 95.7\% & 24 \\
2023 Q4-A\newline (2023-10-02) & 363 & 0.327 & 0.256 & 96.8\% & 26 \\
2023 Q4-B\newline (2023-11-13) & 274 & 0.392 & 0.309 & 82.0\% & 29 \\
\bottomrule

\end{longtable}

\section{Detailed compensation-timing results}
\label{app:timing_detail}

Table~\ref{tab:timing_quarterly_full} reports the integrated Log-CCA timing correlations by target quarter. The first two columns are zero-inclusive outcomes and therefore test whether the score identifies economically material timely relief. The third is conditional on positive eventual relief and asks whether the score predicts the exact waiting-time ordering.

\begin{table}[H]
\centering
\caption{Quarter-level Log-CCA correlations with compensation-timing outcomes.}
\label{tab:timing_quarterly_full}
\small
\begin{tabular}{lrrr}
\toprule
Quarter & Relief within 30 days & Relief-days & Shorter conditional term \\
\midrule
Q1 2022 & 0.561 & 0.551 & 0.187 \\
Q2 2022 & 0.586 & 0.587 & 0.253 \\
Q3 2022 & 0.564 & 0.566 & 0.225 \\
Q4 2022 & 0.583 & 0.585 & 0.203 \\
Q1 2023 & 0.584 & 0.575 & 0.211 \\
Q2 2023 & 0.494 & 0.551 & 0.193 \\
Q3 2023 & 0.479 & 0.525 & 0.209 \\
Q4 2023$^{a}$ & 0.419 & 0.492 & 0.133 \\
\bottomrule

\end{tabular}
\begin{minipage}{0.94\textwidth}\footnotesize
Values are Spearman correlations. $^{a}$Q4 2023 has incomplete observed follow-up and is excluded from complete-quarter medians.
\end{minipage}
\end{table}

Table~\ref{tab:timing_quartiles_full} provides the median integrated-outcome pattern across the seven fully observed quarters. It reinforces the distinction between incidence and exact conditional delay.

\begin{table}[H]
\centering
\caption{Compensation incidence and timing by frozen Log-CCA quartile.}
\label{tab:timing_quartiles_full}
\small
\resizebox{\textwidth}{!}{%
\begin{tabular}{lrrrrr}
\toprule
Quartile & Relations & Positive within 30 days & Positive eventually & Mean delay if positive & Mean days before due \\
\midrule
Q1 lowest & 513 & 57.6\% & 58.5\% & 8.8 & 19.0 \\
Q2 & 512 & 86.2\% & 90.9\% & 12.9 & 35.1 \\
Q3 & 512 & 84.3\% & 91.8\% & 14.9 & 31.1 \\
Q4 highest & 512 & 96.5\% & 98.0\% & 3.3 & 14.9 \\
\bottomrule

\end{tabular}%
}
\end{table}

\begin{figure}[H]
\centering
\includegraphics[width=0.82\textwidth]{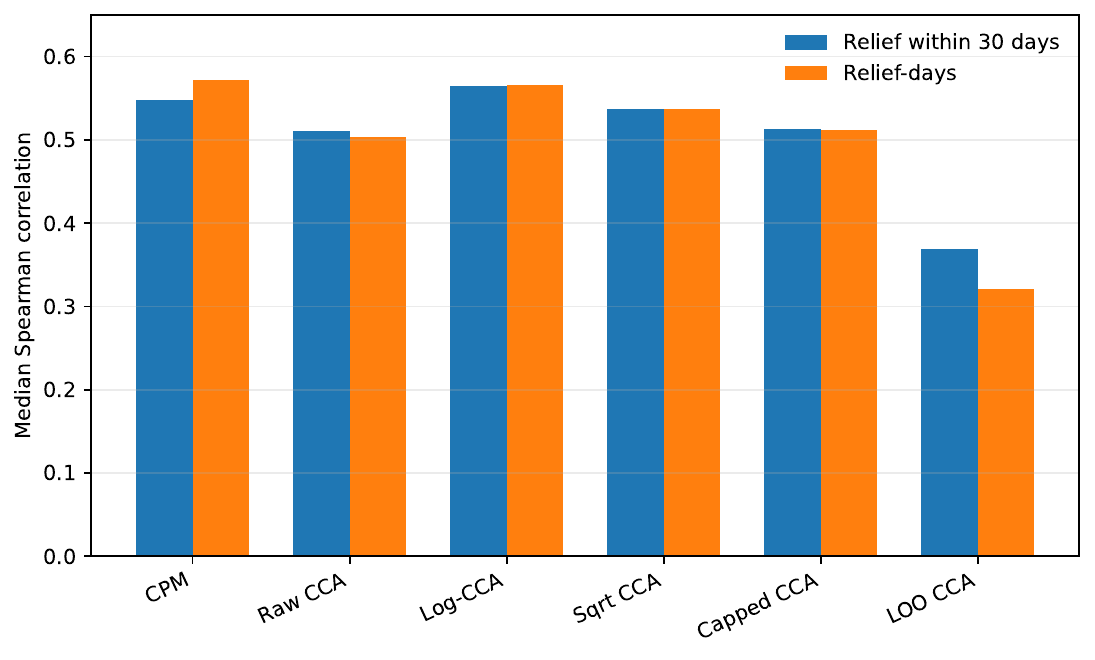}
\caption{Median complete-quarter correlation of formula variants with integrated thirty-day relief and relief-days. CPM is marginally strongest for relief-days, while Log-CCA is strongest for thirty-day relief and provides the most balanced operational specification.}
\label{fig:timing_variants}
\end{figure}

\section{Annual reconciliation with the companion algorithm study}

\begin{table}[H]
\centering
\caption{Annual bridge-adjusted PMR reconciliation in percentage points.}
\label{tab:reconciliation}
\scriptsize
\resizebox{\textwidth}{!}{%
\begin{tabular}{rrrrrrr}
\toprule
Year & Reference cycle & Rerun cycle & Difference & Reference path & Rerun path & Difference \\
\midrule
2012 & 1.3407 & 1.3407 & +0.0000 & 0.9108 & 0.9108 & -0.0000 \\
2013 & 0.0379 & 0.0379 & -0.0000 & 0.0364 & 0.0364 & -0.0000 \\
2014 & 0.2255 & 0.2255 & +0.0000 & 0.7925 & 0.7925 & +0.0000 \\
2015 & 0.2537 & 0.2537 & +0.0000 & 0.3549 & 0.3549 & +0.0000 \\
2016 & 1.2021 & 1.2021 & +0.0000 & 1.3075 & 1.3071 & -0.0004 \\
2017 & 3.8130 & 3.8130 & +0.0000 & 4.3427 & 4.3438 & +0.0011 \\
2018 & 3.4529 & 3.4529 & -0.0000 & 5.0824 & 5.0596 & -0.0228 \\
2019 & 14.1400 & 14.1400 & +0.0000 & 15.8582 & 15.9051 & +0.0469 \\
2020 & 42.7905 & 42.7905 & -0.0000 & 47.7576 & 47.5963 & -0.1613 \\
2021 & 36.7300 & 36.7300 & -0.0000 & 43.4099 & 43.4975 & +0.0876 \\
2022 & 52.7948 & 52.7948 & -0.0000 & 56.4554 & 56.8263 & +0.3709 \\
2023 & 40.7050 & 40.7050 & -0.0000 & 47.5624 & 47.7310 & +0.1686 \\
\bottomrule

\end{tabular}}
\end{table}

The cycle reproduction is exact to the displayed precision. The path differences reflect independent deterministic ordering and reach a maximum of 0.371 percentage points. The bridge and non-reuse identities are exact in both streams.

\section{Reconciliation with the superseded restarted-window analysis}
\label{app:restart_reconciliation}

The preceding release evaluated Q4 2022 through Q4 2023 by restarting each future outcome window. Tables~\ref{tab:restart_direct} and~\ref{tab:restart_selection} compare that superseded construction with the rolling non-reuse stream. Q1--Q3 2022 have no restarted-window counterpart and are therefore absent from the reconciliation.

\begin{table}[H]
\centering
\caption{Direct log-CCA reconciliation: restarted windows versus bridge-correct rolling state.}
\label{tab:restart_direct}
\small
\resizebox{\textwidth}{!}{%
\begin{tabular}{@{}lrrrrrr@{}}
\toprule
& \multicolumn{3}{c}{Spearman} & \multicolumn{3}{c}{Log-Pearson} \\
\cmidrule(lr){2-4}\cmidrule(lr){5-7}
Quarter & Restarted & Bridge-correct & Change & Restarted & Bridge-correct & Change \\
\midrule
Q4 2022 & 0.643 & 0.663 & +0.020 & 0.640 & 0.656 & +0.016 \\
Q1 2023 & 0.631 & 0.647 & +0.016 & 0.616 & 0.632 & +0.015 \\
Q2 2023 & 0.593 & 0.597 & +0.005 & 0.591 & 0.602 & +0.011 \\
Q3 2023 & 0.583 & 0.592 & +0.009 & 0.600 & 0.605 & +0.005 \\
Q4 2023$^{a}$ & 0.549 & 0.529 & -0.020 & 0.534 & 0.491 & -0.043 \\
\bottomrule

\end{tabular}}
\begin{minipage}{0.95\textwidth}\footnotesize
Positive changes indicate a larger coefficient after enforcing physical non-reuse. $^{a}$Q4 2023 has only 39 observed bridge days and is not fully comparable.
\end{minipage}
\end{table}

\begin{table}[H]
\centering
\caption{Buyer-selection reconciliation for frozen log-CCA.}
\label{tab:restart_selection}
\small
\resizebox{\textwidth}{!}{%
\begin{tabular}{@{}lrrrrrr@{}}
\toprule
& \multicolumn{3}{c}{Best relief captured} & \multicolumn{3}{c}{Uplift over mean candidate} \\
\cmidrule(lr){2-4}\cmidrule(lr){5-7}
Quarter & Restarted & Bridge-correct & Change (pp) & Restarted & Bridge-correct & Change (pp) \\
\midrule
Q4 2022 & 91.7\% & 92.5\% & +0.8 & 439.7\% & 420.2\% & -19.5 \\
Q1 2023 & 82.2\% & 68.5\% & -13.6 & 178.0\% & 171.0\% & -7.0 \\
Q2 2023 & 83.5\% & 69.1\% & -14.4 & 183.6\% & 158.6\% & -25.0 \\
Q3 2023 & 95.0\% & 95.2\% & +0.2 & 358.5\% & 348.4\% & -10.0 \\
Q4 2023$^{a}$ & 84.8\% & 86.4\% & +1.6 & 315.8\% & 359.3\% & +43.5 \\
\bottomrule

\end{tabular}}
\begin{minipage}{0.95\textwidth}\footnotesize
The buyer set can change slightly because the bridge-correct stream changes which candidate relations receive positive relief. The restarted results are reported only for transparency and are superseded by the bridge-correct estimates. $^{a}$Q4 2023 is right-censored.
\end{minipage}
\end{table}

The direct association is robust to the correction: all complete overlapping quarters move by less than 0.020 in Spearman correlation. Buyer-level allocation is more sensitive because one early consumption can change the best relation inside a buyer's small candidate set. This is precisely why physical continuity is required even when aggregate rank statistics appear stable.

\section{Deterministic bridge-correct algorithm}

\begin{lstlisting}
for method in {path, cycle}:
    introduced_UIDs = empty set
    residual = empty ledger
    for issue_year y in chronological order:
        if y is the first year:
            introduce invoices issued Jan-Dec y on their issue dates
        else:
            Jan-Feb y invoices already exist only at their carried residual
            introduce Mar-Dec y invoices on their issue dates
        execute causal daily operations on active residual records
        introduce Jan-Feb y+1 as the terminal bridge
        execute causal daily operations through the bridge
        attribute PMR by consumed source fragment and issue cohort
        close unresolved source records from cohort y
        retain only residual Jan-Feb y+1 records for the next phase
        assert original = bridge_consumed + carried_residual
        assert every UID introduced once and consumption <= face value
\end{lstlisting}

For a target forecast window, the analysis does not restart this algorithm. It filters the immutable contribution log by source issue date and execution date. This is what prevents the same bridge amount from becoming available again in the next period.

\section{Reproduction workflow}

\begin{lstlisting}
python src/prepare_atomic_and_phases.py --input data/raw/invoices.csv
c++ -O3 -std=c++17 src/rolling_bridge_engine.cpp \
  -o bin/rolling_bridge_engine
bin/rolling_bridge_engine --method cycle --cycle-cap 8 ...
bin/rolling_bridge_engine --method path ...
python scripts/build_history_atomic.py --history-months 9 ...
python scripts/prepare_origins.py --years 2022 2023 ...
python scripts/analyze_bridge_correct.py --bootstrap 500 ...
python scripts/build_bridge_diagnostics.py ...
python scripts/generate_assets_v4.py ...
python tests/verify_release.py
\end{lstlisting}

Public reproduction verifies aggregate outputs and synthetic accounting tests. Full source-level replay requires authorized access to the pseudonymized atomic ledger and operation contributions.

\end{document}